\documentclass[a4paper,11pt]{article}
\pdfoutput=1
\usepackage{jheppub}
\usepackage{amsmath,amssymb,amsthm,bm}
\usepackage{graphicx}
\usepackage[colorinlistoftodos]{todonotes}
\usepackage[inline]{enumitem}
\definecolor{darkblue}{rgb}{0.1,0.2,0.6}
\definecolor{darkred}{rgb}{0.8,0.1,0.2}
\definecolor{crimson}{RGB}{164,16,52}
\definecolor{darkgreen}{rgb}{0.31,0.62,0.24}
\usepackage{diagbox}
\usepackage{epigraph}
\usepackage{float}
\usepackage{multirow}
\usepackage{tikz}
\usepackage{mathtools}
\usepackage{braket}
\usepackage{soul} % to use \ul
\usepackage[all]{xy}
\usepackage{lipsum}
\usepackage{bbm}
\usepackage[thinlines]{easytable}

\newcommand{\Rom}[1]{\uppercase\expandafter{\romannumeral#1}}

\newcommand{\ex}[1]{\left\langle #1 \right\rangle}

\newcommand{\da}{\dagger}

\newcommand{\pa}{\partial}

\newcommand{\mc}{\mathcal}

\DeclareMathOperator{\sign}{sgn}

\makeatletter
\renewcommand*\env@matrix[1][*\c@MaxMatrixCols c]{%
	\hskip -\arraycolsep
	\let\@ifnextchar\new@ifnextchar
	\array{#1}}
\makeatother

\newtheorem*{claim*}{Claim}

\newcommand{\onlinecite}[1]{\nocite{#1}\citenum{#1}}

\usepackage[normalem]{ulem}

\definecolor{darkred}{rgb}{0.8,0.1,0.2}

\newcommand{\ScalingFactor}{0.8333}

\title{Deep Boundary Perturbations at a Quantum Critical Point}
\author[a,b,c]{Shang Liu}
\affiliation[a]{Institute of Physics,
Chinese Academy of Sciences, Beijing 100910, China}
\affiliation[b]{Department of Physics, California Institute of Technology, Pasadena, California 91125, USA}
\affiliation[c]{Kavli Institute for Theoretical Physics, University of California, Santa Barbara, California 93106, USA}

\emailAdd{sliu.phys@gmail.com}
\abstract{In this work, we explore an unconventional class of problems in the study of (quantum) critical phenomena, termed ``deep boundary criticality''. Traditionally, critical systems are analyzed with two types of perturbations: those uniformly distributed throughout the bulk, which can significantly alter the bulk criticality by triggering a nontrivial bulk renormalization group flow, and those confined to a boundary or subdimensional defect, which affect only the boundary or defect condition. Here, we go beyond this paradigm by studying quantum critical systems with boundary perturbations that decay algebraically (following a power law) into the bulk. By continuously varying the decay exponent, such perturbations can transition between having no effect on the bulk and strongly influencing bulk behavior. We investigate this regime using two prototypical models based on (1+1)D massless Dirac fermions. Through a combination of analytical and numerical approaches, we uncover exotic scaling laws in simple observables and observe qualitative changes in model behavior as the decay exponent varies. }

\begin{document}
\maketitle
\flushbottom
%\tableofcontents

\section{Introduction}
Quantum and classical critical phenomena, which can arise from continuous phase transitions and other contexts, are of fundamental importance in condensed matter and high-energy physics \cite{SachdevQPTBook,CFTBook}. In traditional studies of critical systems and nearby phases, two types of perturbations are typically considered: those uniformly distributed throughout the bulk and those supported on a boundary or subdimensional defect. The first type can significantly affect bulk properties by triggering a nontrivial bulk renormalization group (RG) flow. For instance, even a tiny magnetic field added to the critical Ising model disrupts the long-range correlations, effectively polarizing all spins at low energies. In contrast, the second type of perturbations is not expected to alter the bulk criticality\footnote{We are not aware of a general proof of this claim, but at least for a conformal field theory with a conformally invariant boundary/defect condition, boundary/defect perturbations do not induce any bulk RG flow to all orders in perturbation theory. } but can modify the boundary or defect conditions.

The study of boundary and defect conditions in critical systems, known as boundary or defect criticality, has a long history \cite{CardyBook,Diehl1997Review}, with the famous Kondo problem serving as a classic example \cite{AffleckKondoReview}. This area has attracted renewed attention recently, partly due to connections with gapless symmetry-protected topological (gapless SPT) states. For early developments on this latter topic, see Refs.\,\onlinecite{Grover2012gSPT,Zhang2017gSPT2d,Scaffidi2017gSPT,Parker2018TopLL,Ding2018gSPT2d,Keselman2018GaplessTopSC,Weber2018gSPT2d,Verresen2021gSPT,Ji2020ConformalManifold,Xu2020gSPT3d,Verresen2020TITransitions,Jian2021gSPT2d,Hu2021RandgSPT,Thorngren2021igSPT}. Other significant recent advances include proofs \cite{Friedan2004gThm,Casini2016gThm} and generalizations \cite{Cuomo2022DefectgThm,Casini2023DefectgThm} of the $g$-theorem \cite{AffleckLudwiggThm} on boundary and defect RG flows, the introduction of a new analytic bootstrap approach \cite{Kaviraj2018AnalyticBootstrapBCFT,Zhou2019AnalyticBootstrapBCFT}, the discovery of the extraordinary-log boundary/defect universality class \cite{Metlitski2020ExtraordinaryLog,ParisenToldin2021ExtraordLog,Hu2021ExtraordLog,Metlitski2023ONModelDefect}, and applications of the fuzzy-sphere numerical method \cite{He2023FuzzySphere,Zhu2024DefectFuzzySphere,Zou2024DefectFuzzySphere}, a technique with broader relevance. The list is by no means complete. 

In this work, we go beyond conventional settings to study a distinctive type of critical phenomenon, which we dub \emph{deep boundary criticality}. For concreteness, suppose we have a (1+1)D conformal field theory (CFT) defined on the half-infinite spatial domain $x \geq 0$. We deform the theory by adding the following term to the Lagrangian density: 
\begin{align} 
{\lambda O(\tau,x)}/{x^\alpha}, \end{align} 
where $O(\tau,x)$ is a local operator and $\tau\in\mathbb{R}$ represents imaginary time. This perturbation decays algebraically from the boundary into the bulk, controlled by the parameter $\alpha > 0$. When $\alpha$ is large, so that the perturbation is strongly confined to the boundary, we expect a boundary CFT problem where the bulk criticality remains unaffected. In contrast, for small $\alpha$, particularly in the $\alpha \rightarrow 0$ limit, the perturbation can exert a nontrivial influence on the bulk, provided that $O$ has a sufficiently small scaling dimension $\Delta_{O}$. Power counting suggests that the critical decay exponent separating these scenarios is $\alpha_* = 2-\Delta_{O}$. 
This setup allows us to explore the universal physics that arises when $\alpha<\alpha_*$, $\alpha > \alpha_*$, and $\alpha = \alpha_*$. As a remark, deep boundary perturbations consist of strictly local terms and are distinct from long-range interactions, though they may appear similar.

We restrict our attention to two simple models based on the (1+1)D Dirac fermion CFT. Although these are free fermion models, they exhibit surprisingly nontrivial behavior while offering both analytical and numerical tractability. We uncover exotic scaling laws in simple observables and indeed observe qualitative changes in model behavior as $\alpha$ varies through $\alpha_*$. A more detailed summary of our results is given in the next section. 

Earlier works\footnote{We are grateful to Ferenc Igl\'{o}i, Zohar Komargodski, Chris Herzog, and Siwei Zhong for communications regarding this matter. } have also considered power-law boundary perturbations; see, for example, Refs.\,\onlinecite{Hilhorst1981Inhomogeneous,Igloi1993Review,Gutperle2012BCFTHolography,Bobev2014SupersymmJanus,Horowitz2015HoveringBH,Herzog2019Marginal,Herzog2020Marginal}. However, even the simplest cases of free theories remain not well understood; in particular, the main results presented in this work have not been previously uncovered. 
Field theories with spacetime-dependent couplings but without boundaries or defects have been explored as well, such as in Refs.\,\onlinecite{Dong2012SpacetimeDepCouplings1,Dong2012SpacetimeDepCouplings2} where the couplings exhibit power-law spacetime dependence. 
%Lastly, we emphasize that the models analyzed here feature perfectly local Hamiltonians, distinguishing them from systems with long-range interactions, despite apparent similarities. 

At the end of this introduction section, let us also mention a practical motivation for us to consider such unconventional problems. In recent years, with the advances in quantum technologies, a variety of interesting quantum many-body systems have been realized in quantum simulation platforms. We notice that power-law decaying long-range interactions can be generated in some of those platforms, including trapped ions \cite{Islam2013TrappedIonExpLR,Porras2004TrappedIon} and Rydberg atom arrays. In the former case, the decaying power of the interaction is even continuously tunable. It is then natural to ask the following question: If we quantum simulate a many-body system with a boundary or defect (which is often inevitable in experiments), can the long-range interactions lead to unconventional boundary or defect phenomena? Here, to make life simpler, we choose to not consider long-range bulk interactions which is an active subject of research by itself, but only keep long-range interactions between the boundary and the bulk. If we further fix the boundary degrees of freedom, e.g. by polarizing a boundary spin, the problem reduces to that of deep boundary perturbations. Systems with short-range bulk interactions and deep boundary perturbations may also be experimentally realized at least with superconducting qubits. 

\section{Summary of Results}
We focus on the (1+1)D massless Dirac fermion field theory with a two-component spinor $\psi:=(\psi_+,\psi_-)^{T}$, placed on the half-infinite space $x\geq 0$. Without loss of generality, we impose the boundary condition $\psi_+ + \psi_- =0$ at $x=0$. A more general boundary condition is $\psi_+ + e^{i\gamma}\psi_- =0$, but we can always utilize the chiral U(1) symmetry to set $\gamma=0$. We then add the following \emph{deep boundary perturbation} to the Hamiltonian
\begin{align}
    \Delta H=\int_\delta^\infty dx~\frac{\lambda}{x^\alpha}m(x), 
\end{align}
where $\lambda$ is a coupling constant, $\alpha>0$ controls the decaying of the perturbation, $m(x)$ is a bulk local operator to be specified, and $\delta>0$ is a short-distance cutoff. We choose $m(x)$ to be either of the following two mass operators: 
\begin{align}
    m_Y:=\psi^\da\sigma_y\psi,\quad  m_X:=\psi^\da\sigma_x\psi. 
\end{align}
This produces two models: If $m=m_Y$ ($m=m_X$), the theory will be called Model Y (Model X). 
We note that $m_Y$ and $m_X$ have different symmetry properties. The unperturbed Dirac fermion theory with the chosen boundary condition enjoys a unitary particle-hole symmetry: $\psi\leftrightarrow\psi^\dagger$. While $m_Y$ preserves this symmetry, $m_X$ is antisymmetric. As a side remark, using the chiral U(1) symmetry, one could also map $m_X$ to $m_Y$, at the cost of modifying the boundary condition. This alternative perspective will later save us some work when solving the two models. 

We are interested in universal properties of these two models for different values of $\alpha$ and $\lambda$. On general grounds, given $\Delta_m=1$ in both models, we expect qualitatively different behaviors for $\alpha>1$ and $\alpha<1$. Moreover, $\alpha=1$ with arbitrary $\lambda \neq 0$ marks unconventional renormalization group (RG) fixed points that have been rarely discussed before.

Our main focus is the conceptually most straightforward observable: vacuum expectation value of $m(x)$ in the thermodynamic limit (infinite system size). Just a reminder, $m$ has different meanings in the two models. 
By combining numerical exact diagonalization for large systems and analytical solutions, we are able to determine the leading asymptote of $\ex{m(x)}$ at large $x$ for all $\alpha$ and $\lambda$. In particular, the magnitude of $\lambda$ need not be small\footnote{This statement more precisely means that the dimensionless quantities $\lambda\delta^{1-\alpha}$ and $\lambda(\delta')^{1-\alpha}$ need not be small, where $\delta$ is the short-distance cutoff near the boundary and $\delta'$ is the short-distance cutoff near the operator insertion point $x$. In our numerical calculations, both $\delta$ and $\delta'$ are determined by the lattice constant $a$. }. The results are listed in Table~\ref{tab:LeadingAsymptote}, where we can clearly see the distinction between $\alpha<1$ and $\alpha>1$. 
Note that when $\alpha=1$, $\ex{m(x)}$ does \emph{not} scale as $1/x$ which one may na\"{i}vely conclude from scale invariance. 

\begin{table}
\centering
\begin{TAB}(@)[5pt]{|c|cc|}{|c|c:cc|}% (rows,min,max)[tabcolsep]{columns}{rows}
  & Model Y & Model X  \\
Unperturbed & $\displaystyle\frac{1}{2\pi x}$ & 0\\
$\alpha>1$ & $\displaystyle\frac{1}{2\pi x}$ & $\displaystyle\frac{1}{2\pi x}\tanh\left[\frac{-2\lambda\delta^{1-\alpha}}{\alpha-1}\right]$\\
$0<\alpha\leq 1$  & $\displaystyle
        \frac{-\alpha\lambda\log x}{\pi x^\alpha}$ & $\displaystyle
        \frac{-\alpha\lambda\log x}{\pi x^\alpha}$\\
\end{TAB}
\caption{Leading asymptote of $\ex{m(x)}$ at large $x$. Recall that $m$ is a collective symbol that has different meanings in the two models: $m=m_Y$ ($m=m_X$) in Model Y (Model X). The second row is for the unperturbed theory, while for the next two rows, deep boundary perturbations with $\lambda\neq 0$ are turned on. }
\label{tab:LeadingAsymptote}
\end{table}

For the special value $\alpha=1$, we also obtain very interesting results about higher order corrections. Define
\begin{align}
    \beta:=\begin{cases}
        |2\lambda-1| & \text{for Model Y}\\
        |2\lambda| & \text{for Model X}
    \end{cases}, 
\end{align}
and focusing on the regime $0<\beta<1$, we find the following contribution in the mass expectation value.  
\begin{align}
    \boxed{
    \alpha=1:\quad\ex{m(x)}\supset \frac{\rm const}{x^{1+\beta}}
    }
\end{align}
Moreover, this is the most significant component other than the $\log(x)/x$-scaling and $1/x$-scaling terms. In other words, when $\alpha=1$, 
\begin{align}
    \ex{m(x)}= \frac{-\lambda\log(x)}{\pi x}+\frac{C}{x}+\frac{C'}{x^{1+\beta}}+\text{higher orders}. 
\end{align}

As a byproduct of our analytical approach for $\alpha=1$, we are also able to make a nontrivial prediction about the boundary operator spectrum in this case\footnote{We are grateful to the anonymous referee for pointing this out. }. We find that for $\lambda<1/2$ and $\lambda>1/2$, there should exist a boundary operator with the scaling dimension $n(1/2-\lambda)$ and $n(1/2+\lambda)$, respectively, for any positive integer $n$. 

We shall make one comment regarding the relation to a previous work. As pointed out in Ref.\,\onlinecite{Herzog2019Marginal}, for the marginal case $\alpha=1$ and ignoring the effect of short-distance cutoff, our deep boundary perturbation can actually be mapped to a uniform mass term in two-dimensional anti-de Sitter (AdS\textsubscript{2}) spacetime. However, since this mapping is not an isometry (a Weyl transformation), we believe its effect on the short-distance cutoff is tricky: A spacetime uniform cutoff in our flat spacetime will be mapped to a nonuniform one. Possibly due to this reason, there is a mismatch between our result and that in Ref.\,\onlinecite{Herzog2019Marginal}: the logarithm in our one-point function $\ex{m(x)}$ is absent in the previous work. The precise relation between this work and the AdS\textsubscript{2} field theory deserves further exploration in the future. 

The rest of this paper is organized as follows. In Section~\ref{sec:ModelY}, we present detailed analytical analysis and numerical results for Model Y, including a discussion about topological edge state not summarized above. In Section~\ref{sec:ModelX}, we elaborate on results about Model X, with an emphasis on its difference from Model Y. We conclude and discuss future directions in Section~\ref{sec:Conclusion}. Additional technical details are provided in the appendices.

\section{Model Y: Dirac Fermions with a Particle-Hole Symmetric Perturbation}\label{sec:ModelY}
\subsection{Unperturbed Theory}\label{sec:UnperturbedTheory}
The unperturbed theory that we will consider is the (1+1)D massless Dirac fermion theory described by the following Euclidean Lagrangian density. 
\begin{align}
    \mc L_0=\psi^\da_+(\pa_\tau+i\pa_x)\psi_+ + \psi^\da_-(\pa_\tau-i\pa_x)\psi_-, 
    \label{eq:UnperturbedDirac}
\end{align}
where $\psi_\pm(\tau,x)$ is the left/right-moving complex fermion field. 
We place the theory on the half-infinite space manifold $x\geq 0$, and impose the boundary condition
\begin{align}
    \psi_+(\tau,0)+\psi_-(\tau,0)=0. 
    \label{eq:DiracBdryCondition}
\end{align}
The Lagrangian density \eqref{eq:UnperturbedDirac} and the boundary condition \eqref{eq:DiracBdryCondition} both preserve the particle-hole symmetry $\psi_\pm\leftrightarrow \psi_\pm^\dagger$. 

Correlation functions in this free fermion theory can be computed using the doubling trick. For all $x\geq 0$, define $\psi_+(\tau,-x)=-\psi_-(\tau,x)$ which respects the boundary condition \eqref{eq:DiracBdryCondition}. Then, we can map the original nonchiral theory with a spatial boundary to a chiral theory without spatial boundary: 
\begin{align}
    S_0&=\int_{x\geq 0}d\tau dx~\mc L_0=\int_{x\in\mathbb{R}} d\tau dx~ \psi^\da_+(\pa_\tau+i\pa_x)\psi_+. 
\end{align}
The fermion two-point correlation function in this chiral theory can be obtained from the Schwinger-Dyson equation, and takes the following form. 
\begin{align}
    \ex{\psi_+(z_1)\psi^\da_+(z_2)}=\frac{1}{2\pi (z_1-z_2)}, 
\end{align}
where the complex coordinate $z=\tau+ix$ has been used and $\psi_+(z_i)\equiv\psi_+(\tau_i,x_i)$. 

The field theory \eqref{eq:UnperturbedDirac} with the boundary condition \eqref{eq:DiracBdryCondition} has a simple lattice regularization which enables numerical simulations. This lattice theory lives on a half-infinite chain with sites labeled by $j=1,2,3,\cdots$. There is a single flavor of complex fermions hopping on this chain, described by the following Hamiltonian. 
\begin{align}
    H_{{\rm latt},0}=t\sum_{j=1}^\infty (c_j^\da c_{j+1}+\textrm{h.c.}), 
    \label{eq:LatticeDiracHamiltonian}
\end{align}
where $t>0$ and $c_j$ is the fermion annihilation operator on site $j$. Let $a$ be the lattice constant, or equivalently, the spatial coordinate of the $j$-th site be $x_j=ja$. We take $t=1/(2a)$ so that the fermi velocity equals to $1$. The lattice and continuum fermion operators can be identified as follows. 
\begin{align}
    \frac{c_{x/a}}{\sqrt{a}}\simeq e^{ik_F x}\psi_+(x)+e^{-ik_F x}\psi_-(x), 
    \label{eq:LattContOpCorresp}
\end{align}
where $k_F=\pi/(2a)$ is the fermi momentum. For completeness, we derive this lattice-continuum correspondence in Appendix \ref{app:LatticeDiracFermions}. We observe that this lattice model also has an exact particle-hole symmetry $c_j\leftrightarrow(-1)^j c_j^\dagger$ which corresponds to the aforementioned particle-hole symmetry in the continuum.

\subsection{Deep Boundary Perturbation}
The deep boundary perturbation that we will study in this section takes the following form. 
\begin{align}
    \Delta\mc L=\frac{\lambda}{x^\alpha}m_Y(\tau,x),\quad
    m_Y:=-i(\psi^\da_+\psi_- - \psi^\da_-\psi_+). 
    \label{eq:DeepmYperturbation}
\end{align}
We may cut off $\Delta\mc L$ at a small distance $\delta$ near the boundary, so that the total action takes the form
\begin{align}
    S=\int_{x\geq 0} d\tau dx~\mc L_0 +\int_{x\geq \delta}d\tau dx~\Delta\mc L. 
    \label{eq:ModelYAction}
\end{align}
This is referred to as Model Y throughout this paper. 
Note that $m_Y$ is a mass operator for the Dirac fermions; that is, the fermions will become massive if $m_Y$ is uniformly added to the Lagrangian. An alternative mass operator is $m_X:=\psi^\da_+\psi_- + \psi^\da_-\psi_+$, and in the next section, we will discuss the effect of a deep boundary $m_X$ perturbation. Under the particle-hole transformation, $(m_Y,m_X)\leftrightarrow(m_Y,-m_X)$. 

The lattice counterpart of $m_Y$ is as follows. 
\begin{align}
    &m_Y(x)\simeq \frac{M_{Y,x/a}}{a}, \\
    &M_{Y,j}:=\frac{1}{2}(\tilde M_{Y,j-1/2}+\tilde M_{Y,j+1/2}), \\
    &\tilde M_{Y,j+1/2}:=\frac{1}{2}(-1)^j\left(c^\da_j c_{j+1}+c^\da_{j+1}c_j \right). 
\end{align}
See Appendix \ref{app:LatticeDiracFermions} for the derivation. This correspondence enables us to numerically simulate the perturbed theory. More precisely, we can add the following term to the unperturbed lattice Hamiltonian in \eqref{eq:LatticeDiracHamiltonian}. 
\begin{align}
    \Delta H_{\rm latt}=\sum_{j=2}^\infty \frac{\lambda a^{1-\alpha}}{j^\alpha}M_{Y,j}. 
    \label{eq:LatticeDeltaHModelY}
\end{align}
Here, the sum over $j$ starts from $j=2$ because $M_{Y,j}$ is only well-defined for $j\geq 2$. In actual numerical computations, we need to consider a finite lattice with $N$ sites, the $j$ sum is then also constrained by the upper bound $j\leq N-1$. All numerical results presented in this paper have $N=2\times 10^4$. 

Here comes a technical remark about our numerical simulation. If we wish to precisely reproduce the field theory model \eqref{eq:ModelYAction}, what we should really do is the following: (1) Choose the summation lower bound in \eqref{eq:LatticeDeltaHModelY} to be $j_0$ and then (2) take the limit $a\rightarrow 0$, $N\rightarrow\infty$, $j_0\rightarrow\infty$ while fixing $L=Na$ and $\delta=j_0 a$. Moreover, in order to ensure that the unperturbed fermion dispersion is strictly linear, as discussed in Appendix \ref{app:LatticeDiracFermions}, we need to also (3) manually enforce a fixed cutoff $\Lambda$ on the allowed single-particle momenta while taking the aforementioned limit. However, this set of procedures is too complicated to implement in practice, so we have instead fixed $a$, $N$, $j_0$ and did not impose any other UV cutoff other than the intrinsic lattice regularization. As a result (see Appendix \ref{app:LatticeContinuumMatchingSubtlety}), the effective continuum Lagrangian we simulate will also contain many higher derivative terms which are irrelevant by power-counting. We are expecting that those terms will not affect the long-distance physics we care about, especially given that our models are quadratic. The good match between field theory and numerics (to be presented below) also verifies this expectation. 

The scaling dimension of $m_Y$ is $1$. Therefore, 
just by na\"{i}ve power counting, we expect the perturbation to be relevant (irrelevant) for the bulk criticality when $\alpha<1$ ($\alpha>1$), and $\alpha=1$ is the marginal case. This is reflected in the energy spectrum of the system on a finite space. Let $L$ be the length of the space. In the absence of any perturbation, the finite-size energy gap is of the order $1/L$. After we turn on the perturbation, the Dirac fermions acquire a spatially varying mass that is everywhere no less than $\lambda/L^\alpha$. Hence, if $\alpha<1$, there should be an energy gap of the order $\lambda/L^\alpha$. On the contrary, if $\alpha\geq 1$, we expect the energy gap to be still of the order $1/L$. We have verified these claims numerically. 

In the rest of this section, we investigate the effect of the deep boundary perturbation on the mass expectation value $\ex{m_Y(x)}$. 

\subsection{Mass Expectation Value: Leading Asymptote}
\begin{figure*}
    \centering
    \includegraphics[scale=\ScalingFactor]{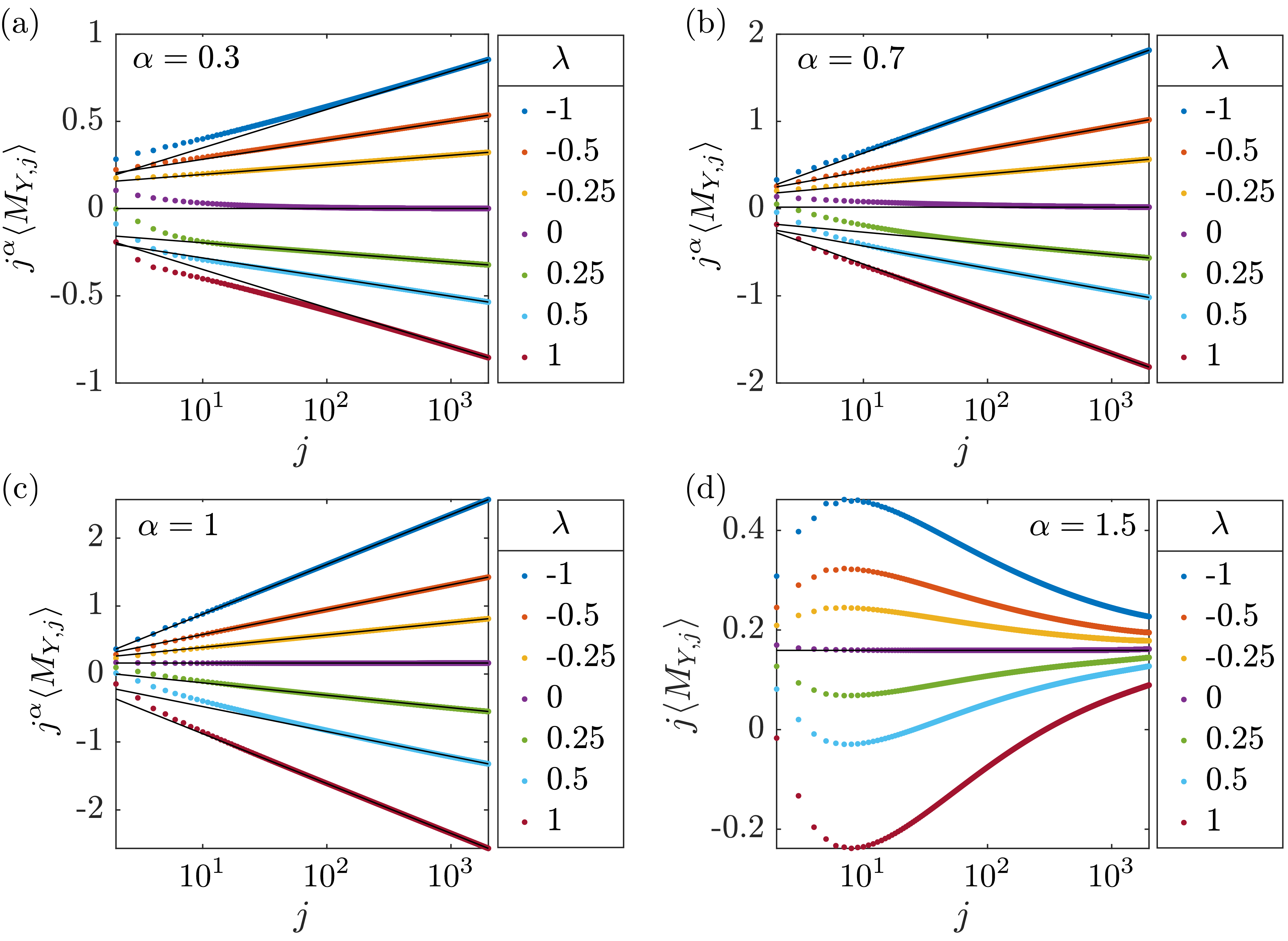}
    \caption{Results on $\ex{M_{Y,j}}\simeq a\ex{m_Y(ja)}$ in Model Y (Eq.\,\ref{eq:DeepmYperturbation}). (a)-(c) are examples with $\alpha\leq 1$, where we plot $j^\alpha \ex{M_{Y,j}}$ against $j$ in log scale. We use the lattice constant $a$ as the length unit, i.e. $a=1$. The total number of sites is $N=20000$, but we only show results for 10\% of the sites to suppress finite-size effect. 
    Dots with different colors are the numerical data, while black solid lines are fittings to the curves $j^\alpha \ex{M_{Y,j}} = -(\alpha\lambda/\pi)\log j+{\rm const}$. (d) contains examples with $\alpha>1$, where we plot $j \ex{M_{Y,j}}$ against $j$ in log scale. The black solid line is the constant line $j \ex{M_{Y,j}}=1/(2\pi)$. 
    }
    \label{fig:YMassVev}
\end{figure*}

With the deep boundary perturbation in \eqref{eq:DeepmYperturbation}, the asymptotic form of $\ex{m_Y(x)}$ for large $x$ is found to be
\begin{align}
    \ex{m_Y(x)}\sim
    \begin{cases}
        \displaystyle
        -\frac{\alpha\lambda\log x}{\pi x^\alpha} & (\alpha\leq 1)\\
        \displaystyle\frac{1}{2\pi x} & (\alpha>1)
    \end{cases}. 
    \label{eq:mYVevLeading}
\end{align}
Note that the symbol $\sim$ is used here to denote the leading term in an asymptotic expansion with its precise multiplicative factor. 
We obtain this result by combining numerical simulations and analytical solutions. The numerical results are given in Fig.\,\ref{fig:YMassVev}, matching very well with the above expressions. Our analytical approaches will be elaborated below in this subsection. We would like to emphasize that, although the asymptotic expressions presented above are at most linear in $\lambda$, they are supposed to be nonperturbative. In other words, for any fixed $\lambda$ that is not necessarily small, these expressions should correctly capture the leading behavior of $\ex{m_Y(x)}$ as $x\rightarrow\infty$. Our numerical simulations are also not restricted to small values of $\lambda$. 

\iffalse
Moreover, for $\alpha=1$ and generic $\lambda$ ($\lambda\notin \mathbb{Z}+1/2$), we have also found interesting higher-order corrections as follows. 
\begin{align}
    \ex{m_Y(x)}\supset \frac{1}{x}\left[ c_0+c_1\left(\frac{a}{x} \right)^{|2\lambda-1|}+\cdots \right]
\end{align}
The second term has an anomalous scaling dimension and is most significant when $\lambda$ approaches $1/2$. 
In the rest of this subsection, we elaborate on numerical results and analytical solutions that lead to the above conclusions.  
\fi

\subsubsection{Irrelevant Case: \texorpdfstring{$\alpha>1$}{alpha>1}}\label{sec:ModelYIrrelCase}
In the absence of any perturbation, $m_Y$ already has a nonzero vacuum expectation value due to the boundary condition we choose. Using the doubling trick discussed in Section \ref{sec:UnperturbedTheory}, one can check that
\begin{align}
    \ex{m_Y(x)}_0=\frac{1}{2\pi x}, 
\end{align}
where the subscript $0$ represents the unperturbed theory. 

Now if we turn on the deep boundary perturbation \eqref{eq:DeepmYperturbation} with $\alpha>1$, since it is irrelevant by power counting, a 
na\"{i}ve expectation will be that $\ex{m_Y(x)}$ approaches $\ex{m_Y(x)}_0$ for large $x$, or more precisely
\begin{align}
    \lim_{x\rightarrow\infty}\ex{m_Y(x)}/\ex{m_Y(x)}_0=1. 
\end{align}
We have indeed confirmed this result numerically, but the above argument is not reliable for the following reason: Although the perturbation is irrelevant \emph{to the bulk}, it may still have a nontrivial effect on the boundary condition. The boundary condition is actually crucial for determining the expectation value of $m_Y$; for example, if we instead took $\psi_++e^{i\gamma}\psi_-=0$ at the spatial boundary, $\ex{m_Y(x)}$ would become $\cos(\gamma)/(2\pi x)$. Therefore, we need to carefully examine the effect of $\Delta\mc L$ on the boundary condition. 

We will take an RG approach, which is nothing but a scale transformation for a free fermion field theory. We write the action as $S=S_0+\Delta S$ where $S_0$ is the unperturbed theory and 
\begin{align}
    \Delta S=\int_{x\geq\delta} d\tau dx~\frac{\lambda}{x^\alpha}m(\tau,x). 
\end{align}
Here, for later convenience, we allow $m$ to represent either $m_Y$ (as considered here) or $m_X$. In fact, $m$ can even be a linear combination of these two. To implement a scale transformation, we define $x'=x/b$, $\tau'=\tau/b$, $\psi_\pm'(\tau',x')=b^{1/2}\psi_{\pm}(\tau,x)$ for some constant $b>1$. Rewrite $S$ in terms of the primed coordinates and fields, and finally remove the primes by renaming, we find that $S_0$ is invariant, while $\Delta S$ changes to 
\begin{align}
    \Delta S'=\int_{x\geq \delta/b}d\tau dx~\frac{\lambda'}{x^\alpha} m(\tau,x), 
\end{align}
where $\lambda'=\lambda b^{1-\alpha}$. 
To deal with the change in the short-distance cutoff, we utilize the bulk-to-boundary operator expansion: 
\begin{align}
    m(\tau,x)=\sum_{k=0}^\infty C_k x^{\Delta_k-1}\mc O_k(\tau), 
\end{align}
where $\{\mc O_k\}$ is a set of independent local operators living on the boundary with scaling dimensions $\{\Delta_k\}$. We assume that (1) $\mc O_0=1$ is the identity operator, (2) $\mc O_1(\tau)=\,:\psi_+^\da\psi_+:(\tau,0)$ with the normal-ordering defined by 
\begin{align}
    \psi^\da_+(z_1)\psi_+(z_2)=\,:\psi^\da_+(z_1)\psi_+(z_2):+\frac{1}{2\pi(z_1-z_2)}, 
\end{align}
and (3) $\Delta_{k\geq 2}>1$. Applying this expansion to the region $b/\delta\leq x\leq \delta$, we get 
\begin{align}
    \Delta S'&=\int_{x\geq \delta}d\tau dx~\frac{\lambda'}{x^\alpha} m(\tau,x)+\sum_k C_k\int d\tau~\lambda'\mc O_k(\tau)\left[\frac{\delta^{\Delta_k-\alpha}-(\delta/b)^{\Delta_k-\alpha}}{\Delta_k-\alpha}\right]. 
\end{align}
In the special case where $\Delta_k=\alpha$, the term in square bracket should be replaced by $\log b$. Now we take the $b\rightarrow \infty$ limit (IR fixed point) under the assumption $\alpha>1$. One can check that 
\begin{align}
    \Delta S'\rightarrow \frac{C_1\lambda\delta^{1-\alpha}}{\alpha-1}\int d\tau~\mc O_1(\tau)+\text{const}. 
    \label{eq:ExactlyMargBdryPert}
\end{align}
We observe that only an exactly marginal boundary perturbation is left. Next, we need to understand the implication on the mass expectation value. For convenience, we use an enriched notation $\ex{m(x)}[S,\delta']$ to denote the vacuum expectation value of $m(x)$ under the action $S$ and the UV cutoff $\delta'$ at the operator insertion point $x$. Note that $\delta'$ is a new parameter, not to be confused with $\delta/b$. 
The necessity of $\delta'$ can be seen, for example, by perturbatively computing $\ex{m(x)}$. Our RG analysis implies that 
\begin{align}
    \ex{m(x)}[S_0+\Delta S,\delta']=b^{-1}\ex{m(x/b)}[S_0+\Delta S',\delta'/b]. 
\end{align}
We choose a $b$ such that $x/b\gtrsim \delta$ is satisfied\footnote{If $x/b<\delta$, the bulk-to-boundary operator expansion may no longer be valid. }. Now, if we can ignore $\delta'/b$, we can compute the right-hand side of the above equation using perturbation theory, and since there is no separation of scales, the perturbation series should be well-behaved. The presence of $\delta'/b$ is somewhat annoying, but one can check that at the first order of the perturbative expansion, it only enters through a log divergence, and the resulting large number $\log(b)$ is suppressed by the coefficient $\lambda'\propto b^{1-\alpha}$. We hereby make the crucial assumption that there is no significant divergence that ruins the perturbation series. Then, in the limit of large $x$ which implies large $b$ and small $\lambda'$, only the zeroth order term of the perturbation series remains. This is equivalent to computing $\ex{m(x)}$ only with the exactly marginal boundary term in Eq.\,\ref{eq:ExactlyMargBdryPert}.

In the case of $m=m_Y$, one can check using the doubling trick that $C_1=0$. Hence the deep boundary perturbation indeed has no effect at long distance, and we can expect $\ex{m_Y(x)}$ to approach $\ex{m_Y(x)}_0$. On the other hand, when $m=m_X$, $C_1\neq 0$ and we will see the consequence in the next section.

\subsubsection{Marginal Case: \texorpdfstring{$\alpha=1$}{alpha=1}}\label{sec:mYMarginalCase}
When $\alpha=1$, the deep boundary perturbation is marginal to the bulk, and starts to have a nontrivial effect. 
Given that the full Lagrangian is still quadratic, i.e. the perturbed theory is still a free fermion theory, we can try to exactly compute $\ex{m_Y(x)}$ from single-particle wave functions. 
In the following, we will present some key steps of this calculation and leave further details to Appendix \ref{App:ExactSolnsMarginalCase}. 
Let $\phi(x)=(\phi_+(x),\phi_-(x))^T$ be a single-particle energy eigenstate wave function with energy $E$. $\phi$ satisfies the following time-independent Schr\"{o}dinger equation. 
\begin{align}
    [i\sigma_z \pa_x+f(x)\sigma_y]\phi=E\phi, 
    \label{eq:phiTISEmYCase}
\end{align}
where $f(x)$ is the mass profile: $f(x)=\lambda/x$ for $x>\delta$ and $f(x)=0$ for $0<x<\delta$. $\phi$ is also subject to the boundary condition $\phi_+(0)+\phi_-(0)=0$ by Eq.\,\ref{eq:DiracBdryCondition}. 
This differential equation happens to be solvable. Let
\begin{align}
    u(x):=\frac{1}{\sqrt{2}}
    \begin{pmatrix}
        1 & 1\\
        1 & -1
    \end{pmatrix}
    \phi(x).
\end{align}
\iffalse
The Schr\"{o}dinger equation for $u$ is
\begin{align}
    (i\sigma_x\pa_x-f\sigma_y)u=Eu, 
    \label{eq:uTISEmYCase}
\end{align}
which implies the two components of $u$ satisfy decoupled second-order differential equations: 
\begin{align}
    [-\pa^2_x+(\pm f'+f^2)]u_\pm = E^2 u_\pm. 
    \label{eq:uTISE2ndOrdermYCase}
\end{align}
\fi
The general solution of $u(x)$ for $x>\delta$ and generic $\lambda$ ($\lambda\notin \mathbb{Z}+1/2$) is 
\begin{align}
    u&=Au_{1}+Bu_{2},\\
    u_{1}&:=\sqrt{-\pi E x}
    \begin{pmatrix}
        J_{\lambda-1/2}(-Ex)\\
        iJ_{\lambda+1/2}(-Ex)
    \end{pmatrix},\\
    u_{2}&:=\sqrt{-\pi E x}
    \begin{pmatrix}
        J_{-(\lambda-1/2)}(-Ex)\\
        -iJ_{-(\lambda+1/2)}(-Ex)
    \end{pmatrix}, 
\end{align}
where $J_\nu(z)$ is the standard Bessel function. The coefficients $A$ and $B$, possibly depending on the energy $E$, need to be determined by the boundary condition and the normalization condition $\int dx~u^\da_E(x)u_{E'}(x)=2\pi\delta(E-E')$ where the subscripts indicate energy dependence. Details of this analysis are given in Appendix \ref{App:ExactSolnsMarginalCase}, and we found that \emph{in the $\delta\rightarrow 0$ limit}, the coefficients take a very simple form: 
\begin{align}
   (A,B)=
    \begin{cases}
        (0,1) & (\lambda<1/2)\\
        (1,0) & (\lambda>1/2)
    \end{cases}. 
\end{align}
The $\delta\rightarrow 0$ limit is supposed to be the IR fixed point of our theory, which can be seen by applying scale transformations to the action. Therefore, the leading long-wavelength physics should be captured by this simplifying limit. The case of nonzero $\delta$ will be discussed later in this section. We can try to compute $\ex{m_Y(x)}$ with the above solution by integrating over the contributions of all single-particle states with $E<0$. A UV cutoff is needed to regularize the divergent integral. As a result, we have found that at large $x$, 
\begin{align}
    \ex{m_Y(x)}\sim \frac{-\lambda\log(x)}{\pi x}
\end{align}
for both $\lambda<1/2$ and $\lambda>1/2$. This is a special case of our previous claim \eqref{eq:mYVevLeading}. 

The $\log(x)$ factor in the above result is interesting and somewhat surprising. If we na\"ively perform scale transformations to the path integral expression of $\ex{m_Y(x)}$, without worrying about any UV cutoff, we would conclude that $\ex{m_Y(x)}\propto 1/x$ which is incorrect. To get a clue about the subtle effect of UV cutoff, we may try to compute $\ex{m_Y(x)}$ by a perturbative expansion in $\lambda$, again in the $\delta\rightarrow 0$ limit. The result turns out to be 
\begin{align}
   \ex{m_Y(x)} = \frac{1}{2\pi x} -\frac{\lambda\log(2x/\delta')}{\pi x} + O(\lambda^2). 
\end{align}
The logarithm appears at the first order due to the divergent integral 
\begin{align}
    \int d\tau dy \ex{m_Y(\tau,y)m_Y(0,x)}_{0,{\rm c}},
\end{align}
where the subscript ${\rm c}$ stands for connected correlation function, and $\delta'$ is a short-distance cutoff as the two $m_Y$ insertions approach. 
This perturbation theory result already suggests that at large $x$, $\ex{m_Y(x)}$ may decay slower than $1/x$. We note that in this special case, the first-order perturbation theory happens to reproduce the exact asymptotic result, but \textit{a priori}, we lack a strong reason to trust the perturbation theory even for small values of $\lambda$. For instance, a term proportional to $\lambda^2\log^2(x)/x$ might emerge at second order and dominate for large $x$, or the entire series might resum into a term proportional to $1/x^{1+2\lambda}$. Furthermore, the perturbation series may fail to converge, as is likely the case for $\alpha < 1$, which will be discussed below. 

We have seen that the expectation value of the ``bare'' mass operator $m_Y(x)$ contains UV divergence at the first order of the perturbation theory. One may wonder whether we can introduce some counter term to cancel the divergence and perhaps use RG to resum some higher order corrections as in conventional renormalized perturbation theories. An attempt to address this question is given in Appendix \ref{app:CounterTerm}. 

As a side comment, by analyzing the near-zero-energy behavior of the single-particle wave functions, we can also extract information about the boundary operator spectrum or temporal correlation functions. Since this is not the main focus of our work, we leave a more detailed discussion of this matter to the last paragraph of Appendix \ref{sec:ZeroEnergyStateApp}. The result is that for $\lambda<1/2$ and $\lambda>1/2$, there should exist a boundary operator with the scaling dimension $n(1/2-\lambda)$ and $n(1/2+\lambda)$, respectively, for any positive integer $n$. Something exotic must happen when $\lambda=1/2$ and we leave a careful study of this point to future works. 

\subsubsection{Relevant Case: \texorpdfstring{$\alpha<1$}{alpha<1}}
For $\alpha<1$, unfortunately, we are unable to solve the single-particle Schr\"{o}dinger equation. Nonetheless, we may try directly extracting useful information about the observable $\ex{m_Y(x)}$ that we care about. To this end, we define two observables at the single-particle level: 
\begin{align}
    F(E;x)&:=\phi^\dagger_E(x)\sigma_y\phi_E(x),\\
    G(E;x)&:=\phi^\dagger_E(x)\sigma_x\phi_E(x), 
\end{align}
where $\phi_E(x)$ is the single-particle energy eigenstate wave function with eigenvalue $E$. The many-body observable $\ex{m_Y(x)}$ is nothing but an integral of $F(E;x)$ over $E$, properly normalized and regularized. $F$ and $G$ satisfy the following differential equations. 
\begin{align}
    \pa_x F&=2EG,\label{eq:dxFEquation}\\
    \pa_x G&= -2EF+2f|\phi|^2, \label{eq:dxGEquation}
\end{align}
where again, $f(x)$ is the mass profile: $f(x)=\lambda/x^\alpha$ for $x>\delta$ and $f(x)=0$ for $0<x<\delta$. 

To make progress, let us consider the regime where $|E|$ is much larger than $|f(x)|$. In this case, we expect the effect of $f$ to be negligible, and thus 
\begin{align}
    \phi\approx
    \begin{pmatrix}
        C_+(E)e^{-iEx}\\
        C_-(E)e^{iEx}
    \end{pmatrix}. 
\end{align}
The conservation of probability current\footnote{Let $\Phi(x,t)$ be a general time-dependent wave function. When $f=0$, the Schr{\"{o}}dinger equation implies $\pa_t|\Phi|^2=\pa_x\left( \Phi^\da\sigma_z\Phi \right)$. Hence, for a steady state $\Phi(x,t)=\phi(x) e^{-iEt}$, we should have $\phi^\da\sigma_z\phi=0$. } implies $|C_+|=|C_-|$. If we assume that the wave function normalization is determined by its plane wave components as is true in the $\alpha=1$ case, we can further conclude $|C_+|=|C_-|=1$. Eqs.\,\ref{eq:dxFEquation} and \ref{eq:dxGEquation} are now solvable by plugging in $|\phi|^2\approx 2$, and we have found
\begin{align}
    F_{\rm approx}(E;x)&=\left[ 2i\lambda x^{1-\alpha}e^{2iEx} \mathrm E_\alpha(2iEx) + {\rm c.c.} \right]+C(E)\cos[2Ex-\theta(E)].
    \label{eq:FApproxSol}
\end{align}
Here, $C\geq 0$ and $\theta\in\mathbb{R}$ are integration constants, $\mathrm E_\alpha$ is the exponential integral 
\begin{align}
    \mathrm E_\alpha(z):=\int_1^\infty dt~\frac{e^{-zt}}{t^\alpha}, 
\end{align}
and the subscript in $F_{\rm approx}$ indicates that the solution is only an approximate one. The expansion parameter that we have implicitly used in Eq.\,\ref{eq:FApproxSol} is $|f(x)/E|$ which as mentioned above is assumed to be small. This can be seen from Eq.\,\ref{eq:dxGEquation}: Since the second term on the right-hand side contains an extra factor of $f$ compared to the first term, by plugging in the leading-order approximation $|\phi|^2\approx2$, we should be able to solve $F$ and $G$ up to one higher order. The expansion parameter can be made more explicit by considering the $x\rightarrow\infty$ limit with $E$ fixed, which implies $|f(x)/E|\rightarrow 0$. In this limit, the first term of $F_{\rm approx}(E;x)$ reduces to $2\lambda/(Ex^\alpha)=2f(x)/E$. 

We propose that the many-body observable $\ex{m_Y(x)}$ can be computed as follows. 
\begin{align}
    \ex{m_Y(x)}\approx \int_{-\infty}^{0} \frac{dE}{2\pi}~F_{\rm approx}(E;x)(e^{E/\Lambda}-e^{E/E0}), 
\end{align}
where $\Lambda>0$ is a UV cutoff, and $E_0=c |\lambda|/x^\alpha$ with some constant $c>0$ is an IR cutoff. Recall that our approximate solution only holds when $|E|$ is much larger than $|f(x)|=|\lambda|/x^\alpha$. Moreover, when $|E|$ is much smaller than $|f(x)|$, we expect the single-particle wave function to have a very small amplitude at the location $x$, which is because the local energy gap $\sim 2\lambda/x^\alpha$ gives a localization length $\sim x^\alpha/(2\lambda) \ll x$ to the wave function. These considerations motivate the introduction of the IR cutoff. The next question is how to evaluate this integral with unknown $C(E)$ and $\theta(E)$. We claim that the contribution from the cosine part in $F_{\rm approx}$ decays faster than $1/x^\alpha$ at large $x$. To see this, let us single out this part of the integral: 
\begin{align}
    \mc I(x)=\int_{-\infty}^{0}\frac{dE}{2\pi} ~C(E)\cos[2Ex-\theta(E)](e^{E/\Lambda}-e^{E/E0}). 
\end{align}
If we multiply by $x^\alpha$, and define $z=Ex^\alpha$, we have
\begin{align}
    x^\alpha\mc I(x)=
    \int_{-\infty}^{0} \frac{dz}{2\pi}~
    &C(z/x^\alpha)\cos[2zx^{1-\alpha}-\theta(z/x^\alpha)][e^{z/(\Lambda x^\alpha)}-e^{z/(c|\lambda|)}]. 
\end{align}
When $x$ is large, $2zx^{1-\alpha}$ in the cosine varies violently as a function of $z$. On the other hand, the $z$ dependence of all other elements in the integrand either becomes milder or does not change as $x$ goes to infinity. 
We thus expect $\lim_{x\rightarrow\infty} x^\alpha \mc I(x)=0$, just like the Fourier mode of any regular function goes to zero at infinite frequency. We have established that the cosine part of $F_{\rm approx}$ contributes a sub-$ 1/x^\alpha$-scaling term to $\ex{m_Y(x)}$. The remaining part of the integral can actually be done analytically, and the result consists of some hypergeometric functions. Expanding the result at large $x$, one can find a contribution that decays slower: 
\begin{align}
    \ex{m_Y(x)}\sim -\frac{\alpha\lambda\log(x)}{\pi x^\alpha}, 
\end{align}
which is exactly the result we claimed at the beginning of this section. 

It may be interesting to compare with the result from the perturbative expansion in $\lambda$: 
\begin{align}
    \ex{m_Y(x)} = \frac{1}{2\pi x} -\frac{\lambda}{\pi x^\alpha}[\log(x)+{\rm const}] + O(\lambda^2). 
\end{align}
We see that the first-order term correctly predicts the $\log(x)/x^\alpha$ dependence, but the overall factor is wrong. 

\subsection{Mass Expectation Value: Higher-Order Corrections for \texorpdfstring{$\alpha=1$}{alpha=1}}
\begin{figure*}
    \centering
    \includegraphics[scale=\ScalingFactor]{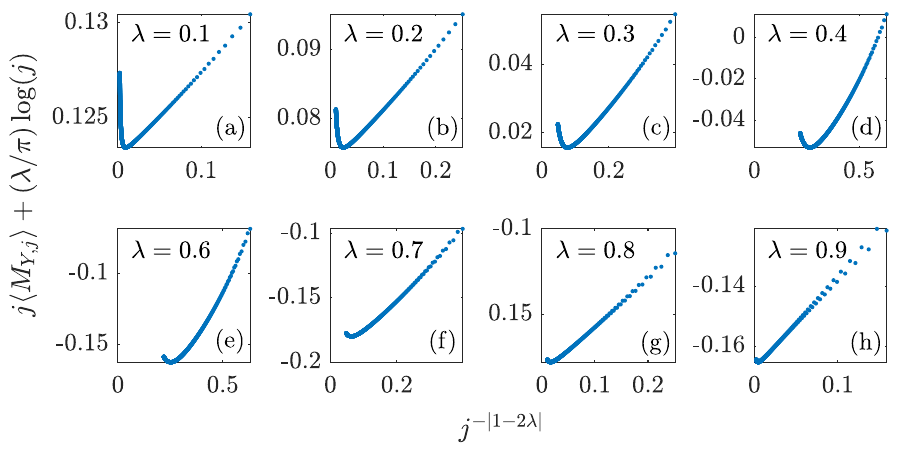}
    \caption{$j\ex{M_{Y,j}}+(\lambda/\pi)\log(j)$ plotted against $j^{-|1-2\lambda|}$ for a few different $\lambda$ in Model Y. Again the total system size is $N=20000$, but data for $j<10$ and $j\gtrsim N/10$ are not shown. }
    \label{fig:YMassVevSub}
\end{figure*}

In our discussion about the exact solution for $\alpha=1$, we saw that the $A,B$ coefficients have a sudden change at $\lambda=1/2$. However, the leading asymptote of $\ex{m_Y(x)}$ does not have any special feature near $\lambda=1/2$. It is then natural to ask whether this special parameter hosts any interesting physical effect at all. As it turns out, there is indeed some interesting phenomenon buried in higher-order corrections to $\ex{m_Y(x)}$, which we now explain. 

For simplicity, we restrict our discussion to $\lambda\in(0,1/2)\cup(1/2,1)$. If we consider $\delta>0$ instead of the $\delta\rightarrow 0$ limit, the coefficients $A,B$ defined in Section \ref{sec:mYMarginalCase} will receive corrections. Let $\zeta:= -E\delta>0$, and $\beta:=|2\lambda-1|$. The corrections are of the order $\zeta^{\beta}$ or higher. As a result, $\ex{m_Y(x)}$ will also receive corrections, including a term that scales as $1/x^{1+\beta}$; this is plausible just by counting the power of $\delta$. In Appendix \ref{Sec:mYNonzerodeltaApp}, we give a more careful analysis of the corrections to $\ex{m_Y(x)}$, in particular the possible divergences in the integrals which may alter the power counting. The conclusions include (1) our previous result on the leading asymptote still holds, (2) there likely exists a $1/x^{1+\beta}$-scaling term, which is the most significant term other than the $\log(x)/x$- and $1/x$-scaling components in the range of $\lambda$ we considered, and (3) there can be many other faster-decaying terms such as one that scales as $1/x^{1+2\beta}$. 

If our prediction is correct, we should have
\begin{align}
    x\left(\ex{m_Y(x)}+\frac{\lambda\log(x)}{\pi x}\right)= C+C'x^{-\beta}+\cdots
    \label{eq:mYHigherOrderCheck}
\end{align}
In Fig.\,\ref{fig:YMassVevSub}, we plotted the left-hand side of the above equation against $x^{-\beta}$ for several values of $\lambda$, using their appropriate lattice counterparts. We see that for most values of $\lambda$ (Panels a-c and f-h), the curves are quite linear at large $x$ (small $x^{-\beta}$) until the finite-size effect blows up as $x$ gets too large. On the other hand, for $\lambda=0.4$ or $0.6$ ($\beta=0.2$), the curve is not so straight. We guess this is because we are unable to achieve very small values of $x^{-\beta}$, so that higher-order corrections are still in effect. Recall there is also a term scaling as $x^{-2\beta}$ on the right-hand side of Eq.\,\ref{eq:mYHigherOrderCheck} which could bent the curve. In fact, the curves in Panels d and e would become quite linear if we changed the horizontal axis to $j^{-2\beta}$, confirming our conjecture. 

\subsection{Topological Edge State}
There is an interesting connection to SPT physics which we shall briefly discuss. If we gap out the Dirac fermion theory by adding the uniform mass term $\lambda m_Y$ to the Lagrangian density (namely $\alpha=0$), the two signs of $\lambda$ correspond to distinct SPT phases protected by the particle-hole symmetry and the charge U(1) symmetry. For the boundary condition we choose, the $\lambda>0$ phase has an edge zero mode while the $\lambda<0$ phase does not. We wonder what is the fate of this topological edge mode in the presence of the symmetric deep boundary perturbation, i.e. in Model Y. 

The $E=0$ energy eigenstate is solvable for arbitrary $\alpha$ and $\lambda$. After incorporating the boundary condition, we have found that 
\begin{align}
    u_{E=0}\propto
    \begin{pmatrix}
        0\\
        \exp\left(-\frac{\lambda }{1-\alpha}x^{1-\alpha}\right)
    \end{pmatrix}
    \quad (\alpha\neq 1). 
\end{align}
For $\alpha=1$, we have instead $u_{E=0}\propto (0,x^{-\lambda})^T$. We observe that for all $\alpha<1$, including $\alpha=0$, the solution is normalizable (unnormalizable) for $\lambda>0$ ($\lambda\leq 0$), and hence a zero-energy bound state exists (does not exist). However, for $\alpha>1$, the solution is not normalizable regardless of the sign of $\lambda$. The $\alpha=1$ case is most interesting: There is a valid bound state for $\lambda>1/2$ but not for $\lambda\leq 1/2$. See Appendix \ref{sec:ZeroEnergyStateApp} for details. 

Notice that $u_{E=0}$ has no contribution to $\ex{m_Y(x)}$ because it has a vanishing component. Therefore, whether or not the bound state is occupied does not affect our previous results.

\section{Model X: Dirac Fermions with a Particle-Hole Antisymmetric Perturbation}\label{sec:ModelX}
\subsection{Deep Boundary Perturbation}
In this section, we will consider another deep boundary perturbation: 
\begin{align}
    \Delta\mc L=\frac{\lambda}{x^\alpha}m_X(\tau,x),\quad
    m_X:=\psi^\da_+\psi_- +\psi^\da_-\psi_+. 
    \label{eq:DeepmXperturbation}
\end{align}
Again, we will cut off $\Delta\mc L$ at a small distance $\delta$ near the boundary, i.e. the change of the action is $\int_{x\geq\delta} d\tau dx~\Delta\mc L$. This is referred to as Model X throughout this paper. As we mentioned previously, $m_X$ is antisymmetric under the particle-hole transformation, which in particular implies that $\ex{m_X(x)}_0=0$ in the unperturbed theory. 

The lattice counterpart of $m_X$ is 
\begin{align}
    &m_X(x)\simeq \frac{M_{X,x/a}}{a}, \\
    &M_{X,j+1/2}:=\frac{1}{2}(\tilde M_{X,j}+\tilde M_{X,j+1}), \\
    &\tilde M_{X,j}:=(-1)^j c^\da_j c_j. 
    \label{eq:LatticemX}
\end{align}
See Appendix~\ref{app:LatticeDiracFermions} for details. The corresponding perturbation to the lattice Hamiltonian is then 
\begin{align}
    \Delta H_{\rm latt}=\sum_{r=r_{\rm min}}^\infty\frac{\lambda a^{1-\alpha}}{r^\alpha}M_{X,r},  
    \label{eq:ModelXLatticeHamiltonian}
\end{align}
where $r\in\mathbb{Z}+1/2$. If not otherwise specified, the $r$ summation starts from $r_{\rm min}=3/2$, the least possible value. 

In the rest of this section, we explore the effect of the deep boundary perturbation on the mass expectation value $\ex{m_X(x)}$. 

\subsection{Mass Expectation Value}
\begin{figure}
    \centering
    \includegraphics[scale=\ScalingFactor]{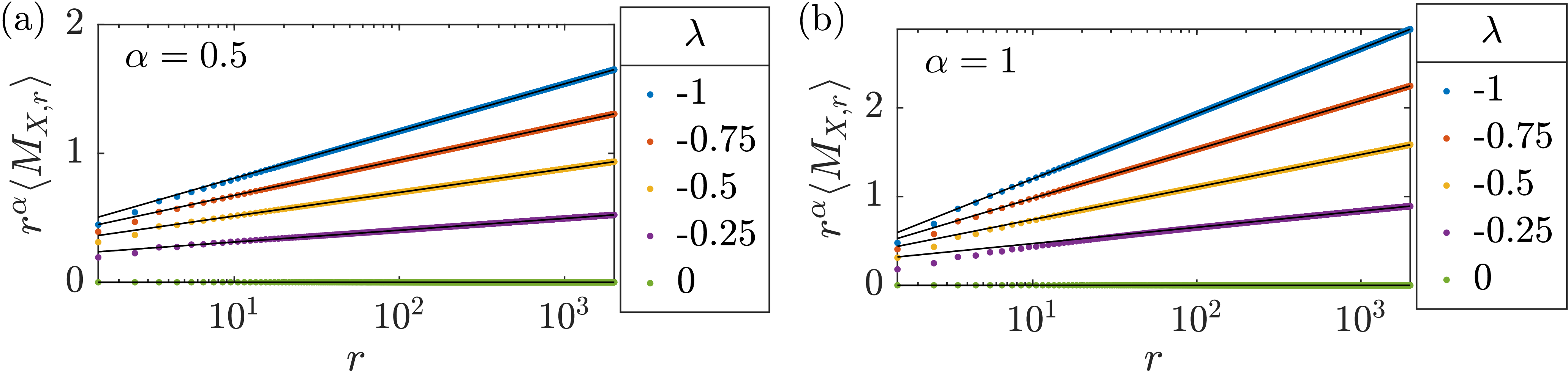}
    \caption{$r^\alpha \ex{M_{X,r}}$ against $r$ in log scale for Model X, where $r\in\mathbb{Z}+1/2$. Panels a and b are for $\alpha=0.5$ and $\alpha=1$, respectively. 
    As before, the lattice constant $a=1$ is the length unit. The total number of sites is $N=20000$, but we only show results for 10\% of the sites to suppress finite-size effect. 
    Dots with different colors are the numerical data, while black solid lines are fittings to the curves $r^\alpha \ex{M_{X,r}} = -(\alpha\lambda/\pi)\log r+{\rm const}$. Note that we restrict to $\lambda\leq0$ because results for $\lambda>0$ are related by the particle-hole transformation. }
    \label{fig:XMassVev_RelMarg}
\end{figure}

\begin{figure}
    \centering
    \includegraphics[scale=\ScalingFactor]{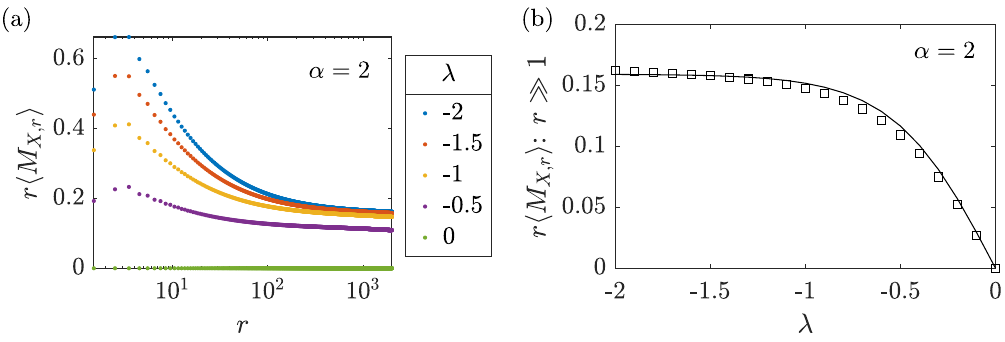}
    \caption{Results on $\ex{M_{X,r}}$ for $\alpha=2$ in Model X. In panel a, we plot $r\ex{M_{X,r}}$ against $r$ for a few values of $\lambda$. In panel b, we fix $r=2000.5\approx N/10$, and plot $r\ex{M_{X,r}}$ against $\lambda$. Squares are the numerical result, while the solid line is the theoretical prediction. }
    \label{fig:XMassVev_Irrel}
\end{figure}

We found the following leading asymptote of $\ex{m_X(x)}$ for large $x$ and infinite system size. 
\begin{align}
    \ex{m_X(x)}\sim
    \begin{cases}
        \displaystyle
        -\frac{\alpha\lambda\log x}{\pi x^\alpha} & (\alpha\leq 1)\\
        \displaystyle\frac{1}{2\pi x}\tanh\left[\frac{-2\lambda\delta^{1-\alpha}}{\alpha-1}\right] & (\alpha>1)
    \end{cases}. 
    \label{eq:mXVevLeading}
\end{align}
Numerical tests are given in Figs.\,\ref{fig:XMassVev_RelMarg} and \ref{fig:XMassVev_Irrel}. In the following, we will explain the analytical derivation of this result, emphasizing on the difference from the previous Model Y case. We will also explain how we estimate $\delta$ for the lattice model. 

\subsubsection{Irrelevant Case: \texorpdfstring{$\alpha>1$}{alpha>1}}
When $\alpha>1$, we may analyze the effect of the deep boundary perturbation with an RG approach as detailed in Section \ref{sec:ModelYIrrelCase}. Using the notations there, it turns out that the leading asymptote of $\ex{m(x)}$ can be computed at the following fixed point: 
\begin{align}
    S=S_0+\Delta S_*,\quad \Delta S_*= \frac{C_1\lambda\delta^{1-\alpha}}{\alpha-1}\int d\tau~\mc O_1(\tau),
    \label{eq:FixedPointActionIrrel}
\end{align}
where $\mc O_1(\tau)=\,:\psi_+^\da\psi_+:(\tau,0)$ is an exactly marginal boundary perturbation. A shortcut to obtain $\Delta S_*$ is applying the bulk-to-boundary operator expansion to the original perturbation $\Delta S$ while retaining only the terms proportional to $\mc O_1$. 
For $m=m_X$, one can check that $C_1=-2$, hence $\Delta S_*$ is nontrivial. To compute $\ex{m_X(x)}$, we need to understand what $\Delta S_*$ does to the boundary condition, and we will take a single-particle quantum mechanics approach. 

Consider the following single-particle spinor Hamiltonian. 
\begin{align}
    &h=i\sigma_z\pa_x+V(x)\sigma_x,\\
    &V(x)=\begin{cases}
        0 & \text{(I)}~0<x<\epsilon/2\\
        -\frac{1}{2}\eta/\epsilon & \text{(II)}~\epsilon/2<x<3\epsilon/2 \\
        0 & \text{(III)}~3\epsilon/2<x
    \end{cases}. 
\end{align}
If we take $\eta=C_1\lambda\delta^{1-\alpha}/(\alpha-1)$, namely the coefficient of $\mc O_1(\tau)$, and then take $\epsilon\rightarrow 0$, this Hamiltonian is equivalent to the fixed point action \eqref{eq:FixedPointActionIrrel}. The solutions to the wave function $(\phi_+,\phi_-)^T$ in these three regions are the following: 
\begin{align}
    &\text{Region I: }
    \begin{pmatrix}
        e^{-iEx}\\
        -e^{iEx}
    \end{pmatrix},\\
    &\text{Region II: }
    Ae^{\lambda_+ x}\begin{pmatrix}
        1\\
        \frac{i\lambda_+ - E}{\eta/(2\epsilon)}
    \end{pmatrix}+
    Be^{\lambda_- x}\begin{pmatrix}
        1\\
        \frac{i\lambda_- - E}{\eta/(2\epsilon)}
    \end{pmatrix},\\
    &\text{Region III: }
    \begin{pmatrix}
        Ce^{-iEx}\\
        De^{iEx}
    \end{pmatrix},  
\end{align}
where $\lambda_\pm=\pm\sqrt{[\eta/(2\epsilon)]^2-E^2}$. Imposing the continuity condition while taking the $\epsilon\rightarrow 0$ limit, we get 
\begin{align}
    e^{i\theta}:=\frac{D}{C}=\frac{-\cosh(\eta/2)+i\sinh(\eta/2)}{\cosh(\eta/2)+i\sinh(\eta/2)}. 
\end{align}
The effective boundary condition at $x=0$ then becomes $\psi_+(\tau,0)-e^{-i\theta}\psi_-(\tau,0)=0$. One can check that, with this boundary condition, 
\begin{align}
    \ex{m_X(x)}=\frac{\tanh(\eta)}{2\pi x}, 
\end{align}
as previously claimed. 

To test the result numerically, we need an estimate of $\delta$ or $\eta$ for our lattice model. Using the lattice-continuum mapping, the lattice model should correspond to the following perturbation to the continuum action: 
\begin{align}
    \Delta S_{\rm latt}=\int d\tau~\sum_{j=1}^\infty \frac{\lambda a}{x_r^\alpha} m_X(\tau,x_r), 
\end{align}
where $r\in \mathbb{Z}+1/2$, $x_r=ra$, and $a$ is the lattice constant. Applying the bulk-to-boundary operator expansion and collect all terms proportional to $\mc O_1$, we find that
\begin{align}
    \eta_{\rm latt}=-2\lambda a^{1-\alpha}\zeta(\alpha,r_{\rm min}), \quad r_{\rm min}=3/2, 
\end{align}
where $\zeta(\alpha,r):=\sum_{k=0}^\infty(k+r)^{-\alpha}$ is the Hurwitz zeta function. The comparison between theory and numerics is given in Fig.\,\ref{fig:XMassVev_Irrel} for $\alpha=2$, especially its Panel b where we plot the lattice version of $x\ex{m_X(x)}$ for a fixed large $x$ against $\lambda$. The matching looks good\footnote{Although we do not have a rigorous proof, we do expect the matching to become perfect if we were able to do numerics at infinite system size and then take the large $x$ limit. }. 

\subsubsection{Marginal Case: \texorpdfstring{$\alpha=1$}{alpha=1}}
\begin{figure}
    \centering
    \includegraphics[scale=\ScalingFactor]{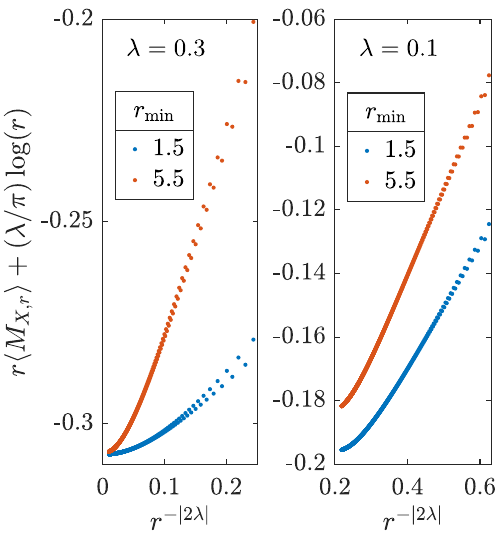}
    \caption{$r\ex{M_{X,r}}+(\lambda/\pi)\log(r)$ against $r^{-|2\lambda|}$ for two choices of $\lambda$ in Model X. $r_{\rm min}$ as defined in Eq.\,\ref{eq:ModelXLatticeHamiltonian} is the lower bound of the summation index $r$. }
    \label{fig:XMassVevSub}
\end{figure}

When $\alpha=1$, we can again exactly solve the single-particle wave functions. There is a trick that reduces some work: If we define
\begin{align}
    (\psi_+,\psi_-)= (\tilde\psi_+,-i\tilde\psi_-), 
\end{align}
then in terms of the tilded fields, the Hamiltonian looks the same as that in Model Y, but the boundary condition gets modified to $\tilde\psi_+ -i\tilde\psi_- =0$. The analysis with this new boundary condition is given in Appendix \ref{App:ExactSolnsMarginalCase} and will not be reproduced here. We end up finding the same leading asymptote for $\ex{m_X}$, namely
\begin{align}
    \ex{m_X(x)}\sim \frac{-\lambda\log(x)}{\pi x}. 
\end{align}

Similar to Model Y, there are interesting higher-order corrections as well. Focusing on $0<\beta:=|2\lambda|<1$, we have found that $\ex{m_X(x)}$ contains a $1/x^{1+\beta}$-scaling term, and it is the most significant component other than the $\log(x)/x$- and $1/x$-scaling terms. To test this term is not easy. In Fig.\,\ref{fig:XMassVevSub}, we plot the lattice version of $x\ex{m_X(x)}+(\lambda/\pi)\log(x)$ against $x^{-\beta}$ for two values of $\lambda$, expecting straight lines. The blue dots are computed using the lattice model in Eq.\,\ref{eq:ModelXLatticeHamiltonian}, and we see that the linearity is good for $\lambda=0.1$ but not satisfactory for $\lambda=0.3$. We then tried terminating the lattice perturbation at a larger value of $r$ (increasing $r_{\rm min}$ in Eq.\,\ref{eq:ModelXLatticeHamiltonian}), effectively increasing $\delta$. This can enhance the $1/x^{1+\beta}$-scaling term because it has a $\delta^{\beta}$ factor, potentially suppressing $\delta$-independent finite-size effect. However, this will also enhance the $1/x^{1+2\beta}$-scaling term more significantly, potentially bending the curve. The result turns out to be good here, as shown by the red dots; the linearity looks much better. On the other hand, the same trick does not seem to help for the previous Model Y case. 

It is interesting to compare again with the result from perturbation theory: 
\begin{align}
    \ex{m_X(x)}= -\frac{\lambda}{\pi x}[\log(2x/\delta')+\log(x/\delta)]+O(\lambda^2), 
\end{align}
where $\delta'$ is a short-distance cutoff near the operator insertion point and $\delta$ is the cutoff near the boundary. The first logarithm leads to the correct $\log(x)/x$-scaling term, while the second logarithm is resummed into those terms with $\lambda$-dependent exponents such as the $1/x^{1+\beta}$-scaling term. Therefore, unlike in Model Y, we would not be able to correctly guess the leading asymptote if we only had the first-order perturbation theory.

\subsubsection{Relevant Case: \texorpdfstring{$\alpha<1$}{alpha<1}}
Finally, for $\alpha<1$, in terms of the tilded fields $\tilde\psi_{\pm}$, the same derivation as that in Model Y can be carried over. The derivation is insensitive to the boundary condition, and we get the same result: 
\begin{align}
    \ex{m_X(x)}\sim \frac{-\alpha\lambda\log(x)}{\pi x^\alpha}. 
\end{align}
A numerical test is shown in Fig.\,\ref{fig:XMassVev_RelMarg}a and we find good agreement.

\section{Conclusion}\label{sec:Conclusion}
In summary, we have investigated the effects of deep boundary perturbations in the (1+1)D massless Dirac fermion theory, uncovering nontrivial behaviors as the decay exponent and coupling constant vary. Through this study, we aim to bring attention to an unconventional class of problems within (quantum) critical phenomena. We hope that our findings will inspire further research and deepen our understanding of boundary effects in critical systems. We suggest a few possible directions for future exploration below. 
\begin{itemize}
    \item There are remaining interesting questions for the two models studied in this work. For example, how do more general correction functions behave? 
    %In particular, how do correlators of the form $\ex{O_1(\tau,x)O_2(0,x)}$ scale for large $\tau$? 
    It is also natural to explore the properties of ground state entanglement entropy. 
    
    \item One can analyze many other models with deep boundary/defect perturbations, especially interacting systems, such as (1+1)D Ising CFT perturbed by a deep boundary magnetic field that explicitly break the Ising $\mathbb{Z}_2$ symmetry. 
    
    \item Another interesting possibility is to turn on multiple deep boundary perturbations and study the interplay between them. 
\end{itemize}

As a final remark, the type of quantum systems studied here may be experimentally realized using superconducting qubits or other quantum simulation platforms. Exploring this possibility is meaningful, as generic strongly interacting critical systems with deep boundary perturbations are difficult to solve exactly using conventional analytical and numerical approaches.

\section*{Acknowledgments}
At various stages of this project, I had the opportunity to share my (preliminary) results with many colleagues, including Jason Alicea, Yimu Bao, Zhen Bi, Ruihua Fan, Yin-Chen He, Wenjie Ji, Ryan Lanzetta, Da-Chuan Lu, Dan Mao, Thomas Scaffidi, David Simmons-Duffin, Ryan Thorngren, Chong Wang, Yifan Wang, Yichen Xu, Weicheng Ye, Yi-Zhuang You, and Yijian Zou. I deeply appreciate their attention and valuable feedback. 
I am also grateful to Ferenc Igl\'{o}i, Zohar Komargodski, Chris Herzog, Siwei Zhong, and Xinan Zhou for reference-related suggestions. 
I acknowledge support from the Gordon and Betty Moore Foundation under Grant No. GBMF8690, the National Science Foundation under Grant No. NSF PHY-1748958, the Simons Foundation under an award to Xie Chen (Award No.  828078), and a start-up grant from the Institute of Physics at Chinese Academy of Sciences.

%\clearpage
\appendix
\section{Lattice Regularization of Dirac Fermions}\label{app:LatticeDiracFermions}
In this appendix, we elaborate on the continuum limit of the lattice fermion model \eqref{eq:LatticeDiracHamiltonian}, and show that it matches with the Dirac fermion theory \eqref{eq:UnperturbedDirac} with the boundary condition \eqref{eq:DiracBdryCondition}. 

\subsection{Periodic Boundary Condition}
Before considering any boundary effect, let us first try to understand the continuum limit of the model on a closed chain with perodic boundary condition. Recall the Hamiltonian takes the following form, 
\begin{align}
    H_{{\rm latt},0}=t\sum_{j} (c_j^\da c_{j+1}+\textrm{h.c.})\quad (t>0). 
    \label{eq:LatticeHamiltonianApp}
\end{align}
In the case of periodic boundary condition, we identify $j$ with $j+N$ where $N$ is the total number of sites. This Hamiltonian can be easily diagonalized by Fourier transform: 
\begin{align}
    c_k=\frac{1}{\sqrt{N}}\sum_j e^{-ik x_j}c_j~ 
    \Leftrightarrow~ c_j=\frac{1}{\sqrt{N}}\sum_k e^{ik x_j} c_k, 
\end{align}
where $x_j=ja$, $a$ is the lattice constant, $k\in (\Delta k) \mathbb{Z}$, $\Delta k=2\pi/(Na)$, and $k$ is identified with $k+2\pi/a$. Be warned that we are somewhat lazy about the notations: We simply use the subscripts ($j$ or $k$) to distinguish operators in the real and momentum spaces. 
In terms of the Fourier modes, 
\begin{align}
    H_{{\rm latt},0}=2t\sum_k \cos(ka)c^\da_k c_k. 
\end{align}
We define the vacuum state as the (many-body) state with all negative-energy modes occupied. Given that $t>0$, the unoccupied modes are in the range $-k_F\leq k \leq k_F$, where $k_F=\pi/(2a)$. The Fermi velocity is $v_F=2ta$. 

The continuum limit more precisely means the following: $N\rightarrow\infty$, $a\rightarrow 0$, $t\rightarrow\infty$, while fixing $L=Na$ that defines the length of the space manifold, and fixing $v_F=2ta$. 
In order to take this limit, we introduce fermion field operators $\psi_\pm(k)$ in the momentum space as follows: 
\begin{align}
    c_{\pm k_F+k}\equiv \sqrt{\frac{\Delta k}{2\pi}}\psi_\pm(k)\quad\text{for}\quad |k|<\Lambda\ll k_F. 
\end{align}
Note that a cutoff scale $\Lambda$ has been introduced. We will fix the value of $\Lambda$ when taking the continuum limit\footnote{One may further take the $\Lambda\rightarrow\infty$ limit on top of this continuum limit. }, and it is assumed that only fermion modes closed to $k=\pm k_F$ are important for the physical quantities that we are interested in. The normalization factor is chosen to reproduce the conventional anticommutation relations as $L\rightarrow\infty$ ($\Delta k=2\pi/L\rightarrow 0$), e.g. $\{ \psi_+(k_1),\psi^\da_+(k_2) \}=(2\pi/\Delta k)\delta_{k_1,k_2}\rightarrow 2\pi\delta(k_1-k_2)$. We further define the real-space fermion field operators $\psi_\pm(x)$ as follows: 
\begin{align}
    \psi_\pm(x)=\sum_{k,|k|<\Lambda}\frac{\Delta k}{2\pi}e^{ikx}\psi_\pm(k)\overset{L\rightarrow\infty}{=}\int_{-\Lambda}^\Lambda\frac{dk}{2\pi}~e^{ikx}\psi_\pm(k), 
    \label{eq:RealSpacepsiDefApp}
\end{align}
which implies, in the continuum limit, 
\begin{align}
    \psi_\pm(k)=\int_{-L/2}^{L/2} dx~e^{-ikx}\psi_\pm(x). 
\end{align}
We can now also establish the lattice-continuum correspondence between real-space operators. Using the Fourier expansion of $c_j$, we have
\begin{align}
    &c_j=\frac{1}{\sqrt{N}}\sum_k e^{ik x_j}c_k\nonumber\\
    &\simeq \frac{1}{\sqrt{N}}\sum_{k,|k|<\Lambda}\left( e^{i(k_F+k)x_j}c_{k_F+k}+ e^{i(-k_F+k)x_j}c_{-k_F+k}\right)\nonumber\\
    &= \sqrt{a}\sum_{k,|k|<\Lambda}\frac{\Delta k}{2\pi}~e^{ikx_j}\left( e^{ik_F x_j}\psi_+(k)+e^{-ik_F x_j}\psi_-(k) \right)\nonumber\\
    &=\sqrt{a}\left( e^{ik_F x_j}\psi_+(x_j)+e^{-ik_F x_j}\psi_-(x_j) \right), 
    \label{eq:LattContOpCorrespApp}
\end{align}
which is nothing but \eqref{eq:LattContOpCorresp} in the main text. 

We can now derive the continuum version of the Hamiltonian. One possible approach is to plug the relation \eqref{eq:LattContOpCorrespApp} into the real-space lattice Hamiltonian. After dropping the rapidly oscillating terms and doing Taylor expansions, we arrive at
\begin{align}
    H_0=v_F\int dx~\left( i\psi^\da_+\pa_x\psi_+ -i\psi^\da_-\pa_x\psi_- \right). 
\end{align}
After setting $v_F=1$, this Hamiltonian corresponds to the standard Dirac Lagrangian \eqref{eq:UnperturbedDirac} in the main text. One can also check that when $N\in 4\mathbb{Z}$, the continuum boundary condition is periodic,  $\psi_\pm(x+L)=\psi_\pm(x)$, while when $N\in 4\mathbb{Z}+2$, the continuum boundary condition becomes antiperiodic, $\psi_\pm(x+L)=-\psi_\pm(x)$. 

We can use \eqref{eq:LattContOpCorrespApp} to derive the lattice-continuum correspondence between other operators as well. For example, 
\begin{align}
    (-1)^j c^\da_j c_j/a \simeq &\left( \psi^\da_+\psi_-+\psi^\da_-\psi_+ \right)(x_j)+e^{2ik_F x_j}\left( \psi^\da_+\psi_++\psi^\da_-\psi_- \right)(x_j). 
\end{align}
Note that the second term on the right-hand side contains an oscillatory factor $e^{2ik_F x_j}=(-1)^j$. Such a term can be removed if we average over two sites, or smear the operator with any function that is smooth in the continuum limit. Therefore, if we define
\begin{align}
    \tilde M_{X,j}&:=(-1)^j c^\da_j c_j,\\
    m_X(x)&:=\left( \psi^\da_+\psi_-+\psi^\da_-\psi_+ \right)(x), 
\end{align}
then
\begin{align}
    M_{X,j+1/2}:=\frac{1}{2}(\tilde M_{X,j}+\tilde M_{X,j+1})\simeq a m_X(x_{j+1/2}). 
    \label{eq:LatticemXApp}
\end{align}
There is actually an implicit subtlety in the above derivation. When we derive \eqref{eq:LattContOpCorrespApp}, we have applied a projection $\mc P$ onto the low-energy subspace defined by $\Lambda$. In other words, \eqref{eq:LattContOpCorrespApp} should be more precisely written as 
\begin{align}
    \mc P c_j\mc P =\sqrt{a}\left( e^{ik_F x_j}\psi_+(x_j)+e^{-ik_F x_j}\psi_-(x_j) \right). 
\end{align}
When we computed the continuum limit of fermion bilinear operators such as the Hamiltonian or $\tilde M_{X,j}$, we should have applied the projection $\mc P$ on the whole operator, instead of on each individual fermion creation/annihilation operator which is what we have done. The net effect is that our results above may be off by (possibly divergent) additive constants. Fortunately, this subtlety does not cause any issue. The Hamiltonian is fine because we do not care about any additive constant there. For \eqref{eq:LatticemXApp}, there is no additive constant because both the lattice and continuum sides have vanishing expectation values on a closed chain: The lattice side can be seen from the lattice translation symmetry, while the continuum side can be seen from the chiral U(1) symmetry whose partial origin is the lattice translation. We assume that the lattice-continuum correspondence of local operators does not depend on the boundary condition. 

Similarly, if we define
\begin{align}
    \tilde M_{Y,j+1/2}&:=\frac{1}{2}(-1)^j\left(c^\da_j c_{j+1}+c^\da_{j+1}c_j \right), \\
m_Y(x)&:=-i\left( \psi^\da_+\psi_- - \psi^\da_-\psi_+ \right)(x), 
\end{align}
then one can check that 
\begin{align}
    M_{Y,j}:=\frac{1}{2}(\tilde M_{Y,j-1/2}+\tilde M_{Y,j+1/2})\simeq a m_Y(x_j). 
    \label{eq:LatticemYApp}
\end{align}

\subsection{Open Boundary Condition}
Now suppose the lattice model lives on an open chain with sites labeled by $j=1,2,3,\cdots,N$, so that the summation in the Hamiltonian \eqref{eq:LatticeHamiltonianApp} runs from $1$ to $N-1$. We need to work out the corresponding boundary condition in the continuum limit. 

The exact single-particle energy eigenstates have the following wave functions: 
\begin{align}
    \phi_k(x_j)=\sqrt{\frac{2}{N+1}}\sin(k x_j),\quad \frac{k(N+1)a}{\pi}=1,2,\cdots,N, 
\end{align}
where $x_j=ja$ as before, and the energy eigenvalues are again $\epsilon_k=2t\cos(ka)$. 
%From this result, we see that it is more natural to define $L=(N+1)a$ as the length of the system. 
These wave functions are nothing but standing waves with nodes at the nonexisting sites $j=0$ and $j=N+1$. 

Given that the single-particle wave functions vanish at $j=0$ and $j=N+1$, we can derive the continuum boundary conditions at both ends by combining \eqref{eq:LattContOpCorrespApp} and the condition $c_0=c_{N+1}=0$. For the left end or $j=0$, we find
\begin{align}
    \psi_+(0)+\psi_-(0)=0, 
\end{align}
as claimed in the main text. For the right end or $j=N+1$, using $k_F=\pi/(2a)$, we find 
\begin{align}
    \psi_+(L)+(-1)^{N+1}\psi_-(L)=0, 
\end{align}
where $L=(N+1)a$. As a sanity check, let us compare the many-body ground state degeneracies of the lattice and continuum models. When $N$ is odd, the continuum theory is equivalent to that of {chiral} fermions living on a {closed} chain with periodic boundary condition, and there is a two-fold ground state degeneracy due to the existence of a zero mode. This is indeed also true in the lattice model. Similarly, when $N$ is even, both the continuum and lattice models have a single ground state. 

\subsection{Comment about our numerical approach}\label{app:LatticeContinuumMatchingSubtlety}
Thus far, we have introduced a precise continuum limit of a lattice model which exactly reproduces the field theory defined in the main text (with a hard momentum cutoff $\Lambda$). In our numerical calculations, however, we did not take any $a\rightarrow 0$ limit, or impose any Hilbert space cutoff $\Lambda$, then how should we interpret these numerical results? 

There is actually an alternative way of thinking about the lattice continuum correspondence. Without taking the $a\rightarrow 0$ limit or imposing the $\Lambda$ cutoff on the allowed momenta, the momentum integration in Eq.\,\ref{eq:RealSpacepsiDefApp} will be from $-\pi/(2a)$ to $\pi/(2a)$. Eq.\,\ref{eq:LattContOpCorrespApp} still holds, but the expressions for the Hamiltonian $H_0$ as well as the mass operators $M_X$ and $M_Y$ will contain higher derivative terms accompanied by higher powers of the lattice constant $a$. We expect that these higher derivative terms do not affect the long-distance behavior of our models, as also verifid by the agreement between numerics and our field theoretic predictions. 

\section{Additional Details about the Exact Solutions for \texorpdfstring{$\alpha=1$}{alpha=1}}\label{App:ExactSolnsMarginalCase}
In this section, we elaborate on the exact solutions to the marginally perturbed Dirac fermion theories, namely the theories with $\alpha=1$. 

\subsection{Model Y}
The continuum Hamiltonian for Model Y is 
\begin{align}
    H&=H_0+\Delta H, \label{eq:HMarginalCaseApp}\\
    H_0&=\int_0^\infty dx~\left( i\psi^\da_+\pa_x\psi_+ -i\psi^\da_-\pa_x\psi_- \right), \\
    \Delta H&=\int_{0}^\infty dx~f(x)m_Y(x). 
\end{align}
For the marginal case concerned here, we can take $f(x)=\lambda/x$ for $x>\delta$ and $f(x)=0$ for $0<x<\delta$. 
The boundary condition is $\psi_+(0)+\psi_-(0)=0$. 

Let $\phi(x)=(\phi_+(x),\phi_-(x))^T$ be a single-particle energy eigenstate wave function with energy $E$. $\phi$ satisfies the following time-independent Schr\"{o}dinger equation. 
\begin{align}
    (i\sigma_z \pa_x+f\sigma_y)\phi=E\phi. 
    \label{eq:phiTISEmYCaseApp}
\end{align}
To solve this equation, it is convenient to define
\begin{align}
    u=
    \begin{pmatrix}
        u_+\\
        u_-
    \end{pmatrix}
    :=U\phi,\quad 
    U:=\frac{1}{\sqrt{2}}
    \begin{pmatrix}
        1 & 1\\
        1 & -1
    \end{pmatrix}=U^\da. 
\end{align}
The wave function $u$ then satisfies
\begin{align}
    (i\sigma_x\pa_x-f\sigma_y)u=Eu. 
    \label{eq:uTISEmYCaseApp}
\end{align}
By applying another copy of the operator $(i\sigma_x\pa_x-f\sigma_y)$ to both sides of the equation, we get 
\begin{align}
    [-\pa^2_x+(\pm f'+f^2)]u_\pm = E^2 u_\pm
    \label{eq:uTISE2ndOrdermYCaseApp}
\end{align}
which can be analytically solved, e.g. by Mathematica. Combining \eqref{eq:uTISE2ndOrdermYCaseApp} and \eqref{eq:uTISEmYCaseApp}, we obtain the following general solution of $u$ for $x>\delta$. 
\begin{align}
    u&=Au_{1}+Bu_{2},\label{eq:uGeneralSol}\\
    u_{1}&:=\sqrt{-\pi E x}
    \begin{pmatrix}
        J_{\lambda-1/2}(-Ex)\\
        iJ_{\lambda+1/2}(-Ex)
    \end{pmatrix},\\
    u_{2}&:=\sqrt{-\pi E x}
    \begin{pmatrix}
        J_{-(\lambda-1/2)}(-Ex)\\
        -iJ_{-(\lambda+1/2)}(-Ex)
    \end{pmatrix}, 
\end{align}
where $J_\nu(z)$ is the standard Bessel function, we assume $\lambda\notin \mathbb{Z}+1/2$ (otherwise we need to use the Neumann function $Y_\nu(z)$ to construct an independent $u_{2}$), and we focus on $E<0$. 

The coefficients $A$ and $B$ need to be determined by the boundary condition and the normalization condition $\int dx~u^\da_E(x)u_{E'}(x)=2\pi\delta(E-E')$ where the subscripts indicate energy dependence of the wave functions. Since we take $f(x)=0$ for $0<x<\delta$, the wave function $\phi_E$ in this region is given by $\phi_E(x)\propto (e^{-iEx},-e^{iEx})^T$ which satisfies the boundary condition $\phi_+(0)+\phi_-(0)=0$. Using the continuity of wave function at $x=\delta$, it follows that 
\begin{align}
    b_Y:=\cos(E\delta)u_+(\delta)+i\sin(E\delta)u_-(\delta)=0. 
\end{align}

One could also try considering a simpler boundary condition: 
\begin{align}
    b_Y':=u_+(\delta)\propto \phi_+(\delta)+\phi_-(\delta)=0. 
\end{align}
This is actually equivalent to terminating the space at $x=\delta$. 
It is hard to say that our lattice model exactly matches with either of the two boundary conditions, so we should only trust results that are insensitive to this choice.

\subsubsection{The \texorpdfstring{$\delta\rightarrow 0$}{delta->0} limit}
Let us first focus on the $\delta\rightarrow 0$ limit. 
Evaluating $b_Y$ for the two solutions $u_1$ and $u_2$ for small $\delta$ gives 
\begin{align}
    b_{Y,1}\sim \delta^\lambda,\quad b_{Y,2}\sim \delta^{1-\lambda}, 
\end{align}
where nonzero normalization factors have been omitted. These imply $u\propto u_2$ for $\lambda<1/2$ while $u\propto u_1$ for $\lambda>1/2$, and exactly the same is predicted by the $b_Y'$ boundary condition. 

In fact, both $u_1$ and $u_2$ are normalized wave functions. To see this, we can utilize the asymptotic expansion of $J_\nu(z)$ for large $z>0$ \cite{Wang1989SpecialFunctions}: 
\begin{align}
    J_\nu(z)&=\sqrt{\frac{2}{\pi z}}\bigg[ \cos\left( 
z-\frac{\nu\pi}{2}-\frac{\pi}{4} \right)+\frac{1-4\nu^2}{8z}\sin\left( z-\frac{\nu\pi}{2}-\frac{\pi}{4} \right)
+O(z^{-2})\bigg]. 
\label{eq:BesselJAsymptoticExpansionApp}
\end{align}
One can check that the plane wave part of $u_i~(i=1,2)$ gives the desired delta function $2\pi\delta(E-E')$ in the integral $\int dx~u^\da_{i,E}(x)u_{i,E'}(x)$, while the remaining part of the integral is finite at $E=E'$. We have therefore identified the single-particle wave functions in the limit of vanishing boundary cutoff: $u=u_2$ ($u=u_1$) for $\lambda<1/2$ ($\lambda>1/2$). 

We can now use the solution to compute the $m_Y$ mass expectation value. Since we are dealing with a free fermion system, this expectation value can be obtained by summing over the contribution of each occupied single-particle mode. We thus have
\begin{align}
    \ex{m_Y(x)}&=\int_{-\infty}^0 \frac{dE}{2\pi}~(-i)(\phi^*_+\phi_- - \phi^*_-\phi_+)_E(x)=\int_{-\infty}^0 \frac{dE}{2\pi}~(-u^\da\sigma_y u)_E(x). 
\end{align}
This integral is actually UV divergent and requires a cutoff. By inserting the exponential cutoff $e^{E/\Lambda}$ into the integrand, a closed form result containing some hypergeometric function can be obtained; let us not spell it out explicitly. By expanding the result for large $x$, we found, for both $\lambda<1/2$ and $\lambda>1/2$, 
\begin{align}
    \ex{m_Y(x)}\sim \frac{-\lambda\log(x)}{\pi x}. 
\end{align}
As also remarked in the main text, if not otherwise noted, the symbol $\sim$ stands for the leading term in an asymptotic expansion with its precise overall factor. 

\subsubsection{Zero-energy states}\label{sec:ZeroEnergyStateApp}
One may wonder whether there are $E=0$ states which have been ignored in the discussions above, and whether including these states will affect our conclusion. We will explain about this issue here. 

Even without solving for the $E=0$ states, we can foresee that they would not affect our previous result on $\ex{m_Y(x)}$. Suppose there exists an $E=0$ state that can be normalized to have unit norm, i.e. a bound state. Either component of this wave function must have an amplitude decaying strictly faster than $1/\sqrt{x}$. As a result, the contribution of this wave function to any fermion bilinear observable decays faster than $1/x$. On the other hand, if there is an $E=0$ state that is not unit normalizable, we may regard it as the limit of a sequence of states at finite system sizes $L$, then as $L\rightarrow\infty$, the unit normalized wave function vanishes everywhere, making no contribution to observables. 

We may directly solve Eq.\,\ref{eq:uTISEmYCaseApp} at $E=0$, which gives 
\begin{align}
    u_{E=0}=
    \begin{pmatrix}
        C_+ x^\lambda\\
        C_- x^{-\lambda}
    \end{pmatrix}\quad(x>\delta). 
    \label{eq:ZeroEnergyStateGeneralSolApp}
\end{align}
The boundary condition (either $b_Y=0$ or $b_Y'=0$) demands $C_+=0$, which is actually true for arbitrary $\delta$, not just in the $\delta\rightarrow 0$ limit. Therefore, $u_{E=0}\propto(0,x^{-\lambda})^T$. Again, we notice that $\lambda=1/2$ is a special value: When $\lambda>1/2$, this is a normalizable bound state solution, but when $\lambda<1/2$, it is not unit normalizable. An interesting observation is that  
\begin{align}
    \lim_{E\rightarrow 0} (-E)^\lambda u_{2,E<0}(x)=
    -\frac{i\sqrt{\pi } 2^{\lambda +1/2}}{\Gamma
   \left(1/2-\lambda\right)}
   \begin{pmatrix}
       0\\
       x^{-\lambda}
   \end{pmatrix}, 
   \label{eq:u2LimitApp}
\end{align}
where $u_{2,E<0}$ is one of the two $E<0$ solutions that we previously found. Recall when $\lambda<1/2$, the boundary condition selects $u_{2,E}$ as the $E<0$ solution, suggesting that $u_{E=0}$ in this case is not an independent solution. In contrast, when $\lambda>1/2$, the boundary condition selects $u_{1,E}$, suggesting that $u_{E=0}$ is an independent solution. We have also verified this conclusion numerically. We examined the finite-size single-particle energy eigenstates for a few values of $\lambda$. There is never a state with exactly $E=0$, so we focus on the highest state with $E<0$ -- the highest occupied state. When $\lambda<1/2$, the wave function of this state appears to match with $u_{2,E<0}$. However, when $\lambda>1/2$, this state behaves very differently from $u_{1,E<0}$. In this case, actually both the highest $E<0$ state and the lowest $E>0$ state look like Eq.\,\ref{eq:ZeroEnergyStateGeneralSolApp}, with both $C_\pm$ nonzero; the $C_+=0$ solution is a superposition of them. The $C_+\neq 0$ component probably comes from $u_{1,E<0}$. Indeed, similar to \eqref{eq:u2LimitApp}, 
\begin{align}
    \lim_{E\rightarrow 0}(-E)^{-\lambda}u_{1,E<0}(x)=\frac{\sqrt{\pi}
   2^{1/2-\lambda }}{\Gamma
   \left(\lambda
   +1/2\right)}
   \begin{pmatrix}
       x^\lambda\\
       0
   \end{pmatrix}. 
   \label{eq:u1LimitApp}
\end{align}

As a side note, the near-zero-energy behavior of single-particle wave functions, such as \eqref{eq:u2LimitApp} and \eqref{eq:u1LimitApp}, actually encode information about the boundary operator spectrum or boundary correlation functions\footnote{We would like to thank the anonymous referee for pointing this out. }. For example, the temopral two-point function $\ex{\psi^\dagger(\tau,x)\psi(0,x)}$ with $\tau>0$ can be computed as 
\begin{align}
    \ex{\psi^\dagger(\tau,x)\psi(0,x)}=\int_{E<0}\frac{dE}{2\pi}~ e^{E\tau}|u_E(x)|^2. 
\end{align}
Now, if the single-particle probability density $|u_E(x)|^2$ contains a term that scales as $|E|^{2\kappa}$ as $E$ approaches zero from below, then the above two-point function will contain a term that scales as $1/\tau^{1+2\kappa}$ for large $\tau$, indicating the existence of a boundary operator with the scaling dimension $1/2+\kappa$. Comparing with \eqref{eq:u2LimitApp} and \eqref{eq:u1LimitApp}, we find that for $\lambda<1/2$, there is a boundary operator with the scaling dimension $1/2-\lambda$, while for $\lambda>1/2$, there is a boundary operator with the scaling dimension $1/2+\lambda$. We can similarly consider temporal correlation functions between fermion operator clusters like the following.  
\begin{align}
    \ex{ \left[ \psi^\dagger(\tau,x_1)\psi^\dagger(\tau,x_2)\cdots \psi^\dagger(\tau,x_n) \right] 
    \left[ \psi(0,x_n)\cdots \psi(0,x_2) \psi(0,x_1) \right]
    }. 
\end{align}
The result is that for $\lambda<1/2$ and $\lambda>1/2$, there should exist a boundary operator with the scaling dimension $n(1/2-\lambda)$ and $n(1/2+\lambda)$, respectively, for any positive integer $n$. 

\subsubsection{\texorpdfstring{$\delta>0$}{delta>0} }\label{Sec:mYNonzerodeltaApp}
When $\delta$ is nonzero, the analysis is more complicated. Let $\chi=B/A$, the condition $b_Y=0$ implies 
\begin{align}
    \chi=\frac{\sin (E\delta )J_{\lambda
   +1/2}(-E\delta)-\cos(E\delta ) J_{\lambda
   -1/2}(-E\delta )}{\sin (E\delta) J_{-(\lambda
   +1/2)}(-E\delta )+\cos (E\delta ) J_{-(\lambda-1/2)}(-E\delta)}. 
\end{align}
Taking into account the wave function normalization, we find
\begin{align}
    (A,B)=\frac{(1,\chi)}{\sqrt{1+2\chi\sin(\pi\lambda)+\chi^2}}. 
\end{align}
Here, as in the previous case, the normalization is determined solely by the plane wave part of the wave function $u(x)$. The expressions for $A$ and $B$ now have complicated energy dependence and we are no longer able to exactly compute any observables. Nonetheless, by analyzing the behavior of the these coefficients at small $E\delta$, we can still make some exact predictions. 

For simplicity, we will assume that $\lambda$ is closed but not equal to $1/2$ which is the interesting special value. More precisely, we consider either $0<\lambda<1/2$ or $1/2<\lambda<1$. 
Let $\zeta=-E\delta>0$. We have 
\begin{align}
    \chi=\zeta^{2\lambda-1}[M_0+O(\zeta^2)],\quad M_0= \frac{-2^{-2\lambda}\Gamma(3/2-\lambda)}{\lambda\Gamma(1/2+\lambda)}. 
\end{align}
The leading-order terms in $A$ and $B$ for small $\zeta$ are the following: 
\begin{align}
    &0<\lambda<1/2:~
    \begin{cases}
        A\approx M_0^{-1}\zeta^{1-2\lambda}&\\
        B\approx 1-M_0^{-1}\sin(\pi\lambda)\zeta^{1-2\lambda}&
    \end{cases}
,\\
    &1/2<\lambda<1:~
    \begin{cases}
        A\approx 1-M_0\sin(\pi\lambda)\zeta^{2\lambda-1}&\\
        B\approx M_0 \zeta^{2\lambda-1}&
    \end{cases},
\end{align}
where an unimportant overall sign has been dropped in the first case. 
Recall that 
\begin{align}
    \ex{m_Y(x)}=\int_{-\infty}^0 \frac{dE}{2\pi}~e^{E/\Lambda}(-u^\da\sigma_y u)_E(x). 
\end{align}
$u^\da\sigma_y u$ can now be expanded as 
\begin{align}
    A^2u_1^\da\sigma_y u_1 +AB(u_1^\da\sigma_y u_2+u_2^\da\sigma_y u_1)+B^2 u_2^\da\sigma_yu_2. 
    \label{eq:mYIntegrandTermsApp}
\end{align}
The $O(\zeta^0)$ terms of the $A,B$ coefficients lead to our previous result of $\ex{m_Y(x)}$ in the $\delta\rightarrow 0$ limit. Our new task now is to analyze the corrections due to nonzero $\zeta$. 

As an example, let us focus on the middle (cross) term in Eq.\,\ref{eq:mYIntegrandTermsApp}. It will contribute the following term to the expectation value of $m_Y$: 
\begin{align}
    \int_{-\infty}^0 {dE}~e^{E/\Lambda}\zeta^{\beta}f(z) ,
\end{align}
where $z:= - Ex$, $\beta=|2\lambda-1|\in(0,1)$, and
\begin{align}
    f(z)= -(2\pi)^{-1}M_0^{\sign(2\lambda-1)} (u_1^\da\sigma_y u_2+u_2^\da\sigma_y u_1). 
\end{align}
By a change of variable, this integral becomes
\begin{align}
    \frac{\delta^\beta}{x^{1+\beta}}\int_0^\infty z^\beta f(z)e^{-z/(\Lambda x)}dz. 
\end{align}
If this integral is convergent in the $\Lambda\rightarrow \infty$ limit, then the result will consist of a $1/x^{1+\beta}$-scaling term plus faster-decaying corrections due to finite $\Lambda$. However, the integral may actually contain UV divergence, so a more careful analysis is needed. We first note that at small $z$, $(u_1^\da\sigma_y u_2+u_2^\da\sigma_y u_1)$ scales as $z^0$, hence the integral has no IR divergence. In fact, one may similarly check that the whole integral for $\ex{m_Y(x)}$ has no IR divergence. Thererfore, we may split the above integral by $\int_0^\infty=\int_0^1+\int_1^\infty$. The integral for $0<z<1$ is finite and gives a $1/x^{1+\beta}$-scaling term plus higher corrections. To analyze the integral in the range $1<z<\infty$, we may expand $f(z)$ for large $z$. Using the known asymptotic expansion of Bessel functions (see Eq.\,\ref{eq:BesselJAsymptoticExpansionApp} which is already enough for our purpose), we observe that the large-$z$ expansion of $f(z)$ only contains the following two types of terms: (1) ${\sin(2z+\theta)}/{z^m}$ for $m=0,1,2,\cdots$, and (2) ${1}/{z^n}$ for $n=1,2,3,\cdots$; note the absence of $n=0$. The same actually holds for all three components of Eq.\,\ref{eq:mYIntegrandTermsApp}. The first type of terms will not lead to any divergence. In fact, $\lim_{\eta\rightarrow 0+}\int_1^\infty dz~z^\gamma \sin(2z+\theta)e^{-\eta z}$ is finite for all $\gamma\in\mathbb{R}$. On the other hand, the second type of terms may lead to divegence if $n-\beta\leq 1$ which is actually only possible for $n=1$ since $0<\beta<1$. In this case, the result consists of $1/x$-scaling, $1/x^{1+\beta}$-scaling, and faster-decaying terms. 

Next, we shall analyze other corrections from nonzero $\zeta$, and the same line of reasoning as above can essentially be carried over. In general, those corrections come from integrals of the following form: 
\begin{align}
    \int_{-\infty}^0 {dE}~e^{E/\Lambda}\zeta^{\Delta}g(z),
\end{align}
where $\Delta\geq\beta$. There is no IR divergence, so the integral can be split as before. The large $z$ behavior of $g(z)$ is also similar: There can only be (1) ${\sin(2z+\theta)}/{z^m}$ for integer $m\geq 0$, and (2) ${1}/{z^n}$ for positive integer $n\geq 1$. The result will consist of $1/x^{1+\Delta}$-scaling or faster-decaying terms, as well as $1/x^n$-scaling terms with positive integer $n\geq 1$. For special values of $\Delta$ such that $n-\Delta=1$ for some $n$, there can also be terms scaling as $\log(x)/x^{1+\Delta}=\log(x)/x^n$, but this is only possible for $n\geq 2$. 

We also note that the zero-energy bound state found in Section \ref{sec:ZeroEnergyStateApp} has no contribution to $\ex{m_Y(x)}$. 

To summarize, with all these efforts, we have found that for $\lambda\in(0,1/2)\cup(1/2,1)$, $\ex{m_Y(x)}$ should contain a term that scales as $1/x^{1+\beta}$ ($\beta=|2\lambda-1|$) at large $x$ unless some magical cancellation happens. Moreover, this is the most significant component other than the $\log(x)/x$- and $1/x$-scaling terms. There also exist higher-order terms such as one that scales as $1/x^{1+2\beta}$. 

\subsection{Model X}
For Model X, it is convenient to define 
\begin{align}
    (\psi_+,\psi_-)= (\tilde\psi_+,-i\tilde\psi_-), 
\end{align}
namely a chiral transformation. We then have 
\begin{align}
    H_0=\tilde H_0,\quad m_X=\tilde m_Y. 
\end{align}
Therefore, in terms of the tilded fields, the Model X Hamiltonian takes the same form as that in Model Y, but the boundary condition becomes $\tilde\psi_+(0)-i\tilde\psi_-(0)=0$. 

For generic $\lambda$, the wave function $\tilde u$ has the same form as that in Eq.\,\ref{eq:uGeneralSol}, with coefficients $\tilde A$ and $\tilde B$. If we turn off the deep boundary perturbation in the region $0<x<\delta$, the boundary condition dictates that 
\begin{align}
    b_X:=&\cos(E\delta)e^{-i\pi/4}(\tilde u_+ + i\tilde u_-)(\delta)+ i\sin(E\delta)e^{i\pi/4}(\tilde u_+ - i\tilde u_-)(\delta)=0. 
\end{align}
Analogous to $b_Y'$ introduced previously, we may also define an alternative boundary condition
\begin{align}
    b_X':=e^{-i\pi/4}(\tilde u_+ + i\tilde u_-)(\delta)=0,  
\end{align}
which is equivalent to terminating the space at $x=\delta$. We are interested in universal properties of the system that does not depend on the choice between $b_X$ and $b_X'$. 

One can check that at small $x$, $b_{X,1}\sim x^\lambda$ and $b_{X,2}\sim x^{-\lambda}$, where nonzero factors have been omitted. In fact, the same holds for $b_{X,1}'$ and $b_{X,2}'$ as well. Hence, in the $\delta\rightarrow 0$ limit, we have $\tilde u=u_1$ for $\lambda>0$ and $\tilde u=u_2$ for $\lambda<0$. Computing the mass expectation value gives
\begin{align}
    \ex{m_X(x)}=\ex{\tilde m_Y(x)}\sim \frac{-\lambda\log(x)}{\pi x}, 
\end{align}
same as in Model Y. 

Next, let us consider the case $\delta>0$. From the small-$x$ expansions of $b_{X,1}$ and $b_{X,2}$, we know that $\tilde\chi:=\tilde B/\tilde A=\tilde M_0\zeta^{2\lambda}+\cdots$ for small $\zeta=-E\delta$ and some nonzero $\tilde M_0$. The same analysis as that for Model Y suggests the existence of a $1/x^{|2\lambda|}$-scaling term in $\ex{m_X(x)}$. When $0<|2\lambda|<1$, this is the most significant component other than the $\log(x)/x$- and $1/x$-scaling terms. 

What about zero-energy states? Demanding either $b_X=0$ or $b_X'=0$ which are the same for $E=0$, the unnormalized zero-energy solution takes the following form, 
\begin{align}
    \tilde u_{E=0}=\begin{pmatrix}
        (x/\delta)^\lambda\\
        i(x/\delta)^{-\lambda}
    \end{pmatrix}. 
\end{align}
This is never normalizable. Moreover, it is a limit of the $E<0$ wave function $\tilde u_{E<0}$; this is simplest to check in the $\delta\rightarrow 0$ limit but holds more generally. Therefore, we believe there is no independent $E=0$ solution. 

\section{Counter Term for the Observable}\label{app:CounterTerm}
In the main text, we have briefly touched on the perturbative expansion of the mass expectation value, where the bare mass operator has been used. One may wonder whether we can define a renormalized operator to remove the UV divergence and whether this could help resum certain higher order corrections by integrating some RG equation. This section serves as an attempt to explore this question. 

We will focus on Model Y with $\alpha=1$. 
%In this case, the RG equation for $\lambda$ should be trivial, because we have a free fermion field theory where the renormalization of the action is equivalent to dimension analysis. 
Since we have a free fermion field theory, it appears reasonable to assume that the renormalization of the action is equivalent to dimension analysis, i.e. the RG equation for $\lambda$ is trivial, though we do not have a rigorous proof. 
However, although the renormalization of the action is trivial, composite operators inserted into correlation functions may still have nontrivial renormalization. 
Recall that perturbation theory gives
\begin{align}
   \ex{m_Y(x)} = \frac{1}{2\pi x} -\frac{\lambda\log(\Lambda x)}{\pi x} + O(\lambda^2), 
\end{align}
where $\Lambda$ is some UV cutoff \footnote{The $\Lambda$ here is not necessarily equal to other uses of the same symbol in this paper. }. The UV divergence at the first order needs to be canceled by some counter term. Here comes two choices: 
\begin{enumerate}
    \item We can define an \emph{additive} counter term such that the renomalized operator $m_{Y,R}(x)$ is given by 
    \begin{align}
        m_{Y,R}(x)=m_{Y}(x)+\delta_m,\quad \delta_m=\frac{\lambda}{\pi x}\log(\Lambda/\mu)+\cdots 
    \end{align}
    where $\mu$ is the renomalization scale. In this case, we are actually not sure whether the term ``renormalization'' is appropriate, as the effect of $\delta_m$ is more similar to normal ordering, but we will use this term anyway. 

    \item Alternatively, we may define a \emph{multiplicative} renomalization factor such that 
    \begin{align}
        m_{Y,R}(x)=m_{Y}(x)Z_m,\quad Z_m=1+2\lambda\log(\Lambda/\mu)+\cdots
    \end{align}
\end{enumerate}
We will try to resum higher order corrections using both approaches to renormalization and compare the results with the exact answer. 

Let us start with the second approach, namely that with a multiplicative renormalization factor. Given that the bare operator has no dependence on the renomalization scale $\mu$, we have
\begin{align}
    0=\frac{d}{d\log\mu}\log\ex{m_Y(x)}&=\frac{d}{d\log\mu}\log[Z_m^{-1}\ex{m_{Y,R}(x)}]\\
    &=2\lambda+\frac{d}{d\log\mu}\log\ex{m_{Y,R}(x)}+O(\lambda^2). 
\end{align}
Let us now integrate $\mu$ from the low-energy scale $x^{-1}$ all the way to $\Lambda$. This gives
\begin{align}
    \log\left[ \frac{\ex{m_{Y,R}(x)}_{\Lambda}}{\ex{m_{Y,R}(x)}_{x^{-1}}} \right]=-2\lambda\log\left( \Lambda x \right)+O(\lambda^2). 
\end{align}
We know that $\ex{m_{Y,R}(x)}_{x^{-1}}=w(\lambda,\Lambda x)/x$ for some dimensionless function $w$ that is finite in the $\Lambda\rightarrow \infty$ limit. Moreover, when $\mu$ is set to be $\Lambda$, the renormalization factor $Z_m$ evaluates to a dimensionless function of $\lambda$, denoted as $z(\lambda)$. We then have
\begin{align}
    z(\lambda) \ex{m_Y(x)}=\frac{w(\lambda,\Lambda x)}{x^{1+2\lambda }\Lambda ^{2\lambda}}[1+O(\lambda^2)]. 
\end{align}
If ignoring the higher order corrections, we have found that $\ex{m_Y(x)}\propto 1/x^{1+2\lambda}$ at large $x$, and this is inconsistent with our exact solution. 

Next, consider the first approach where the counter term is additive. We have
\begin{align}
    0=\frac{d}{d\log\mu}\ex{m_Y(x)}=\frac{d}{d\log\mu}\ex{m_{Y,R}(x)}+\frac{\lambda}{\pi x}+O(\lambda^2). 
\end{align}
Again integrating over $\mu$, we obtain
\begin{align}
    z(\lambda) \ex{m_Y(x)}=\frac{w(\lambda,\Lambda x)}{x}-\frac{\lambda}{\pi x}\log(\Lambda x)+O(\lambda^2). 
\end{align}
The result has nothing new, but is indeed consistent with our exact solution. 

\bibliographystyle{JHEP}
\bibliography{Bib_DeepBoundary.bib}
\end{document}